\documentclass[iop,apj]{emulateapj}

\usepackage{amsmath}
\usepackage{txfonts}
\usepackage{graphicx}
\usepackage{url}
\usepackage{hyperref}

\newcommand{\pd}{\partial}
\newcommand{\ud}{\ensuremath{\mathrm{d}}}

\begin{document}

\title{A New Multi-Dimensional General Relativistic Neutrino
Hydrodynamics Code for Core-Collapse Supernovae
IV. The Neutrino Signal}

\author{Bernhard~M\"uller\altaffilmark{1,2}, Hans-Thomas~Janka\altaffilmark{2}}

\affil{
\altaffilmark{1}{Monash Center for Astrophysics, School of
  Mathematical Sciences, Building 28, Monash University, Victoria
  3800, Australia; \href{mailto:bernhard.mueller@monash.edu}{bernhard.mueller@monash.edu}}
\\
\altaffilmark{2}{Max-Planck-Institut f\"ur Astrophysik,
  Karl-Schwarzschild-Str.~1, D-85748 Garching, Germany;
  \href{mailto:bjmuellr@mpa-garching.mpg.de}{bjmuellr@mpa-garching.mpg.de}, \href{thj@mpa-garching.mpg.de}{thj@mpa-garching.mpg.de}}
}

\begin{abstract}
Considering six general relativistic, two-dimensional (2D) supernova (SN)
explosion models of progenitor stars between 8.1 and 27\,$M_\odot$, we
systematically analyze the properties of the neutrino emission from
core collapse and bounce to the post-explosion phase. The models were
computed with the \textsc{Vertex-CoCoNuT} code, using three-flavor,
energy-dependent neutrino transport in the ray-by-ray-plus
approximation. Our results confirm the close similarity of the mean
energies, $\langle E\rangle$, of $\bar\nu_e$ and heavy-lepton
neutrinos and even their crossing during the accretion phase for stars
with $M\gtrsim 10\,M_\odot$ as observed in previous 1D and 2D
simulations with state-of-the-art neutrino transport. We establish a
roughly linear scaling of $\langle E_{\bar\nu_e}\rangle$ with the
proto-neutron star (PNS) mass, which holds in time as well as for
different progenitors.  Convection inside the PNS affects the neutrino
emission on the 10--20\% level, and accretion continuing beyond the
onset of the explosion prevents the abrupt drop of the neutrino
luminosities seen in artificially exploded 1D models.  We demonstrate
that a wavelet-based time-frequency analysis of SN neutrino signals in
IceCube will offer sensitive diagnostics for the SN core dynamics up
to at least $\sim$10\,kpc distance. Strong, narrow-band signal
modulations indicate quasi-periodic shock sloshing motions
due to the standing accretion shock instability (SASI), and the
frequency evolution of such ``SASI neutrino chirps'' reveals shock
expansion or contraction. The onset of the explosion is accompanied by
a shift of the modulation frequency below 40--50\,Hz, and
post-explosion, episodic accretion downflows will be signaled by
activity intervals stretching over an extended frequency range in the
wavelet spectrogram.
\end{abstract}


\section{Introduction}
\label{sec:intro}

Neutrinos play a fundamental role in core-collapse supernovae: Not
only do they carry away most (several $10^{53} \ \mathrm{erg}$) of the
gravitational binding energy liberated during the collapse of the
inner shells of the progenitor to a proto-neutron star, they are also
the driving agent for the supernova explosion in the most popular
scenario for shock revival, the ``delayed neutrino-driven mechanism''
of \citet{bethe_85}; see
\citet{janka_12,janka_12b,burrows_13,kotake_12b} for current
reviews. Once the explosion has been successfully initiated and
post-shock accretion has ceased, the continuing neutrino emission from
the proto-neutron star drives a baryonic wind that has long been
considered as a site of interesting nucleosynthesis (see, e.g.,
\citealt{arcones_13}).

Moreover, the observation of supernova neutrinos from a future
Galactic event could provide tremendous insights into the dynamics
deep inside the stellar core and serve to constrain unknown particle
physics.  The detection of some two dozen neutrinos from Supernova
SN~1987A \citep{hirata_87,bionta_87,alekseev_87} already confirmed the
basic picture of neutrino emission in core-collapse supernovae,
suggesting the emission of $\sim 3 \times 10^{53} \ \mathrm{erg}$ with
a time-averaged neutrino temperature of $\sim 4 \ \mathrm{MeV}$ from a
neutrinosphere of the order of a few $10 \ \mathrm{km}$ with a total
signal duration of a few seconds (see, e.g.,
\citealt{arnett_89,burrows_90,koshiba_92} for an overview).
The signal from SN~1987A also provided constraints on the mass,
electric charge, magnetic moment and lifetime of neutrinos as well as
indirect constraints on the mass of a hypothetical axion (see
\citealt{burrows_90,keil_97} and references therein).  Due to much
higher event rates, present and next-generation detectors will allow
for the reconstruction of the time-dependent neutrino signal in much
greater detail (see, e.g., \citealt{abe_12,wurm_12} and, for a general
overview, \citealt{scholberg_12}), which could allow far-reaching
conclusions both on the dynamics in the supernova core and open
questions in neutrino physics such as the mass hierarchy
\citep{dighe_00,serpico_12}.

Quantitatively accurate predictions for the neutrino emission from
supernova simulations are an indispensable prerequisite for properly
interpreting the neutrino signal from a prospective Galactic event.
Nowadays, sophisticated methods for the solution of the neutrino
transport problem are available for this purpose. The most advanced
 schemes for neutrino transport in \emph{spherically symmetric} (1D)
neutrino hydrodynamics simulations rely
either on the direct solution of the energy-dependent general
relativistic \citep{yamada_97,liebendoerfer_04} or Newtonian
\citep{mezzacappa_93} Boltzmann equation, or on a variable Eddington
factor method with Boltzmann closure (Newtonian: \citealt{burrows_00},
pseudo-relativistic: \citealt{rampp_02}, general relativistic:
\citealt{mueller_10,roberts_12a}).  These methods
have been used to address different phases of the neutrino
emission from core-collapse supernovae.  Several studies focused
specifically on the neutronization burst
\citep{kachelriess_05,langanke_08} and the rise of the electron
antineutrino and the heavy flavor neutrino luminosity
\citep{serpico_12} around shock breakout (with an emphasis on the
detectable signal). The dependence of the neutrino emission during the
pre-explosion (accretion) phase on the progenitor properties and the
equation of state (EoS) of nuclear matter has been the subject of a
larger number of papers
\citep{thompson_03,liebendoerfer_03,sumiyoshi_05,buras_06_b,fischer_09,oconnor_13,nakazato_13}. \citet{sumiyoshi_08}
and \citet{fischer_09} also cover the case of black hole formation in
failed supernovae.  \citet{lentz_12a,lentz_12b} recently studied the
impact of variations in the microphysics and approximations in the
neutrino transport sector and emphasized the importance of a rigorous
treatment of general relativity, observer corrections, and
energy-exchanges in scattering reactions on electrons and nucleons for
accurate predictions of neutrino luminosities and mean energies.

1D simulations using multi-group Boltzmann or variable Eddington factor
transport have likewise been used to predict the signal from the
Kelvin-Helmholtz cooling phase after the successful initiation of an
explosion
\citep{huedepohl_10,fischer_10,fischer_12,roberts_12a,roberts_12b,martinez_12}.
However, with the exception of electron-capture supernovae
\citep{huedepohl_10,fischer_10}, predictions for the neutrino signal
from the post-explosion phase are currently based on \emph{artificial}
explosion models in which shock revival is achieved by manually
boosting neutrino heating in the gain layer, or even start from
initial models of the proto-neutron star constructed by hand
\citep{roberts_12a,roberts_12b}.

According to our current understanding, neutrino-driven
core-collapse supernovae are inherently \emph{multi-dimensional}
(multi-D).  Hydrodynamic instabilities like buoyancy-driven
convection in the neutrino-heated gain region
\citep{bethe_90,herant_92,herant_94,burrows_95,janka_96,mueller_97}
and the standing accretion shock instability (``SASI'',
\citealp{blondin_03,blondin_06,foglizzo_06,ohnishi_06,
  foglizzo_07,scheck_08,iwakami_08,iwakami_09,fernandez_09,fernandez_10})
were found to play a crucial role in the explosion mechanism, and they
are no less relevant for the neutrino emission. However, predictions
for the neutrino signal from multi-D supernova simulations with a
transport treatment on par with the best available 1D models are still
scarce. Attempts at a rigorous solution of the 3D Boltzmann equation
\citep{kotake_12,radice_13,peres_13} are yet in their infancy.  The
best available studies of multi-D effects on the neutrino emission
therefore either rely on Newtonian 2D multi-angle transport with
considerable compromises in the microphysics and omission
of energy-bin coupling
\citep{ott_08_a,brandt_10} or on multi-group variable Eddington factor
transport using the ``ray-by-ray-plus approximation'' of
\citet{buras_06_a,bruenn_06} in 2D \citep{marek_08,lund_10} and 3D
\citep{tamborra_13}. Predictions for the explosion phase are currently
available only from parameterized models with gray transport and an
excised neutron star core \citep{mueller_e_12,lund_12}. Despite the
different methodologies there is a consensus that the neutrino signal
is considerably affected by multi-D instabilities. Asymmetric
accretion onto the proto-neutron star gives rise to spatial
anisotropies and temporal variations in the neutrino emission. These
spatio-temporal variations show a distinct imprint of the different
hydrodynamic instabilities in the supernova core, with SASI
oscillations leading to fluctuations in the neutrino signal with a
rather well-defined frequency around $100 \, \mathrm{Hz}$, and
convective overturn resulting in more stochastic variations
\citep{tamborra_13}. 
 After the onset of the explosion, the neutrino emission
is typically characterized by large-scale
spatial anisotropies as the accretion flow
onto the proto-neutron star (which usually subsists
for several hundreds of milliseconds after shock revival)
becomes highly asymmetric.

In this paper, we reexamine the neutrino emission in core-collapse
supernovae on the basis of state-of-the-art axisymmetric (2D) general
relativistic simulations with energy-dependent three-flavor neutrino
transport. We consider a wide range of progenitors from the
lowest-mass iron cores, where multi-D effects play a minor role for
the neutrino emission, through more massive models with vigorous
convection and SASI. Different from previous studies 
of neutrino emission in multi-D simulations of the post-bounce phase
\citep{ott_08_a,marek_08,lund_10,brandt_10,tamborra_13}, we explore both
the pre-explosion and the explosion phase. We analyze both the overall
secular evolution of the neutrino emission as well as spatio-temporal
variations on shorter time-scales. In order to connect with future
neutrino observations, we show how the quantitative analysis of the
signal from a prospective Galactic event in a detector like IceCube
\citep{abbasi_11,salathe_12} could provide detailed time-dependent
information about the dynamics in the supernova core.

Our paper is structured as follows: In Section~\ref{sec:numerics}, we
outline the numerical methods and the input physics of our
simulations, including the progenitor models. In
Section~\ref{sec:overview}, we review the secular evolution of the
spherically averaged neutrino luminosities and mean energies.  In
Section~\ref{sec:variations}, we discuss the spatio-temporal
variations in the neutrino emission on the basis of simulated IceCube
signals. By means of a time-frequency analysis using wavelet
transforms, we show that the IceCube signal can reveal the
detailed time-dependence of the SASI sloshing frequency, the rough
evolution of the shock radius, the onset of the explosion, and early
fallback onto the proto-neutron star through newly-formed accretion
funnels. In Section~\ref{sec:summary}, we summarize our findings, examine
their robustness, and discuss further implications for the observation
of a future Galactic supernova in gravitational waves and neutrinos.

\begin{table*}
  \caption{Neutrino physics input
    \label{tab:rates}
  }
  \begin{center}
    \begin{tabular}{cc}
      \hline \hline 
           process  & reference \\
      \hline
      $\nu A \rightleftharpoons \nu A$ & \citet{horowitz_97} (ion-ion correlations) \\
                                       & \citet{langanke_08} (inelastic contribution) \\
      $\nu \ e^\pm \rightleftharpoons \nu \ e^\pm$  &  \citet{mezzacappa_93} \\
      $\nu \ N \rightleftharpoons \nu \ N$          & \citet{burrows_98}\tablenotemark{a} \\
      $\nu_e\ n \rightleftharpoons e^-\ p$      & \citet{burrows_98}\tablenotemark{a} \\
      $\bar{\nu}_e \ p \rightleftharpoons e^+ \ n$  & \citet{burrows_98}\tablenotemark{a} \\
      $\nu_e \ A' \rightleftharpoons e^- \ A$  & \citet{langanke_03} \\
      $\nu\bar{\nu} \rightleftharpoons e^- \ e^+ $ & \citet{bruenn_85,pons_98} \\
      $\nu\bar{\nu} \ NN \rightleftharpoons NN $ & \citet{hannestad_98} \\
      $\nu_{\mu,\tau}\bar{\nu}_{\mu,\tau} \rightleftharpoons \nu_{e}\bar{\nu}_{e} $ 
                                                    & \citet{buras_03} \\
$\stackrel{(-)}{\nu}_{\mu,\tau}\stackrel{(-)}{\nu}_e \rightleftharpoons \stackrel{(-)}{\nu}_{\mu,\tau}\stackrel{(-)}{\nu}_e$ 
                                                    & \citet{buras_03} \\
      \hline \hline 
    \end{tabular}
    \tablenotetext{1}{ Note that these reaction rates account for
      nucleon thermal motions, phase-space blocking, energy transfer
      to the nucleon associated with recoil (``non-conservative'' or
      ``non-isoenergetic'' scattering), and nucleon correlations at
      high densities.  Moreover, we include the quenching of the
      axial-vector coupling at high densities \citep{carter_02},
      correction to the effective nucleon mass \citep{reddy_99}, and
      weak magnetism effects \citep{horowitz_02}. However,
we ignore nucleon potential effects \citep{martinez_12,roberts_12c},
which are of minor importance during the
accretion phase \citep{martinez_12}.}
  \end{center}
\end{table*}

\begin{table*}
  \caption{Model setup
    \label{tab:model_setup}
  }
  \begin{center}
    \begin{tabular}{cccccccc}
      \hline \hline 
            & metallicity & angular    & explosion  & time of                   &      & progenitor & simulation\\
progenitor  & $Z/Z_\odot$ & resolution & obtained   & explosion\tablenotemark{a} & EoS & reference  & reference\\
      \hline
 u8.1   & $10^{-4}$ & $1.4^\circ$ & yes        & $175 \ \mathrm{ms}$ & LS180  & A.~Heger (private communication) & \citet{mueller_12b} \\
z9.6    & $0$       & $1.4^\circ$ & yes        & $125 \ \mathrm{ms}$ & LS220  & A.~Heger (private communication) & \citet{mueller_13}\\
s11.2   & $1$       & $2.8^\circ$ & yes        & $213 \ \mathrm{ms}$ & LS180  & \citet{woosley_02} & \citet{mueller_12a}\\
s15s7b2 & $1$       & $2.8^\circ$ & yes        & $569 \ \mathrm{ms}$ & LS180  & \citet{woosley_95} & \citet{mueller_12a} \\ 
s25   & $1$       & $1.4^\circ$ & no         & ---                 & LS220  & \citet{woosley_02} & \citet{mueller_13} \\
s27   & $1$       & $1.4^\circ$ & yes        & $209 \ \mathrm{ms}$ & LS220  & \citet{woosley_02} & \citet{mueller_12b} \\
      \hline \hline
    \end{tabular}
    \tablenotetext{1}{Defined as the point in time when the average shock radius $\langle r_\mathrm{sh} \rangle$ reaches $400 \ \mathrm{km}$.}
  \end{center}
\end{table*}

\section{Numerical Methods and Model Setup}
\label{sec:numerics}
We study the neutrino emission in six axisymmetric (2D) core-collapse
supernova simulations computed with the general relativistic
neutrino hydrodynamics code \textsc{Vertex-CoCoNuT}
(\citealp{mueller_10}, paper~I in this series).  For full details, the
reader should consult paper~I, since we confine ourselves to a very
brief outline of the code methodology in this section.

\subsection{Hydrodynamics and Gravity}
The hydrodynamics module \textsc{CoCoNuT} is a general relativistic
finite-volume solver using higher-order PPM reconstruction
\citep{colella_84} and the approximate HLLC Riemann solver
\citep{mignone_05_a} modified to avoid odd-even decoupling and the
carbuncle phenomenon at strong shocks \citep{quirk_94}. 
The metric equations are solved in the extended conformal
flatness approximation (xCFC) of \citet{cordero_09}.

\subsection{Neutrino Transport}
The neutrino transport module \textsc{Vertex} integrates the
energy-dependent zeroth and first moment equations for neutrino and
antineutrinos of all flavors using a variable Eddington factor
technique \citep{rampp_02}.  We resort to the ``ray-by-ray-plus''
approximation \citep{buras_06_a} to make the multi-D transport problem
tractable.  In the ray-by-ray-plus approach, the radiation field is
assumed to be axially symmetric around the radial { unit} vector, which
effectively decouple the neutrino transport problem along different
rays through the origin.  However, the lateral advection of neutrinos
with the fluid is included in an operator-split fashion, as is the
lateral neutrino pressure gradient in the optically thick regime. It
should be pointed out that the ray-by-ray-plus approximation does
\emph{not} imply that neutrinos propagate only in the radial
direction, it merely implies that the energy flux vector is radial and
that lateral gradients (except for advection terms) in the moment
equations are neglected.  The ray-by-ray-plus approach allows us to
predict angular variations in the neutrino radiation field at least in
rough qualitative agreement with full multi-angle transport
\citep{ott_08_a,brandt_10}.  Full multi-angle transport smears
out angular variations in the radiation field at larger radii,
especially outside the neutrinosphere. In
order to obtain better quantitative estimates for the angular
variations at large distances from the proto-neutron star, we
therefore reprocess our simulation data
 to obtain observable neutrino fluxes using the method introduced by
\citet{mueller_e_12} as described in Section~\ref{sec:reconstruction}.

The moment equations solved along each ray fully include relativistic
effects \citep{mueller_10} and energy redistribution in ``inelastic''
or ``non-conservative'' scattering reactions. An up-to-date set of
neutrino interactions rates, briefly summarized in
Table~\ref{tab:rates}, has been used for the simulations presented in
this paper. All models were computed with the ``full set'' of
\citet{mueller_12a}.

\subsection{Progenitor Models and Core-Collapse
Simulations} 
We have simulated the collapse and post-bounce evolution of
non-rotating progenitors with $8.1 M_\odot$, $9.6 M_\odot$, $11.2
M_\odot$, $15 M_\odot$, $25 M_\odot$, and $27 M_\odot$. The $8.1
M_\odot$ (u8.1) and $9.6 M_\odot$ (z9.6) stars (A.~Heger, private
communication) are progenitors close to the lower mass limit for iron
core formation at metallicities of $Z=10^{-4}$ and $Z=0$, respectively.
These stars
explode very quickly with little help from convection and the SASI.
Three progenitors (s11.2, s25, s27) have been taken from
\citet{woosley_02} and serve as examples for a relatively fast but
rather weak convectively-dominated explosion (s11.2), a fast
SASI-dominated explosion (s27), and a SASI-dominated model
(s25), which fails to explode until more than 
$600 \, \mathrm{ms}$ after bounce and shows little promise
for a successful explosion at later times. Model s15s7b2 from \citet{woosley_95} illustrates the case of a
late explosion. The dynamics of these models has already been
discussed in much detail elsewhere
\citep{mueller_12a,mueller_12b,mueller_13,janka_12b}.

The core-collapse simulations have been conducted using the equation
of state of \citet{lattimer_91} with a bulk incompressibility modulus
of $K=220 \, \mathrm{MeV}$ (LS220) for models z9.6, s25, and s27, and
with $K=180 \, \mathrm{MeV}$ (LS180) for u8.1, s11.2, and s15s7b2. For
the neutron star masses encountered in these two cases, LS220 and
LS180 yield very similar results because of very similar proto-neutron
star radii as discussed in \citet{mueller_13}, and the incompatibility
of LS180 with the observed maximum neutron star mass of $\approx 2
M_\odot$ \citep{demorest_10} is therefore of no immediate concern.
We expect only a weak dependence of the model dynamics and the
  neutrino signal on the choice of bulk incompressibility ($K=220
  \, \mathrm{MeV}$ vs.\ $K=180 \, \mathrm{MeV}$) for these two EoS's.

Models s11.2 and s15s7sb2 have been simulated with a reduced
angular resolution of $2.8^\circ$ (64 zones) instead
of $1.4^\circ$ (128 zones). The setup of the six simulations is summarized in
Table~\ref{tab:model_setup}, including references for more
details on the model evolution and dynamics.

\begin{figure*}
\plotone{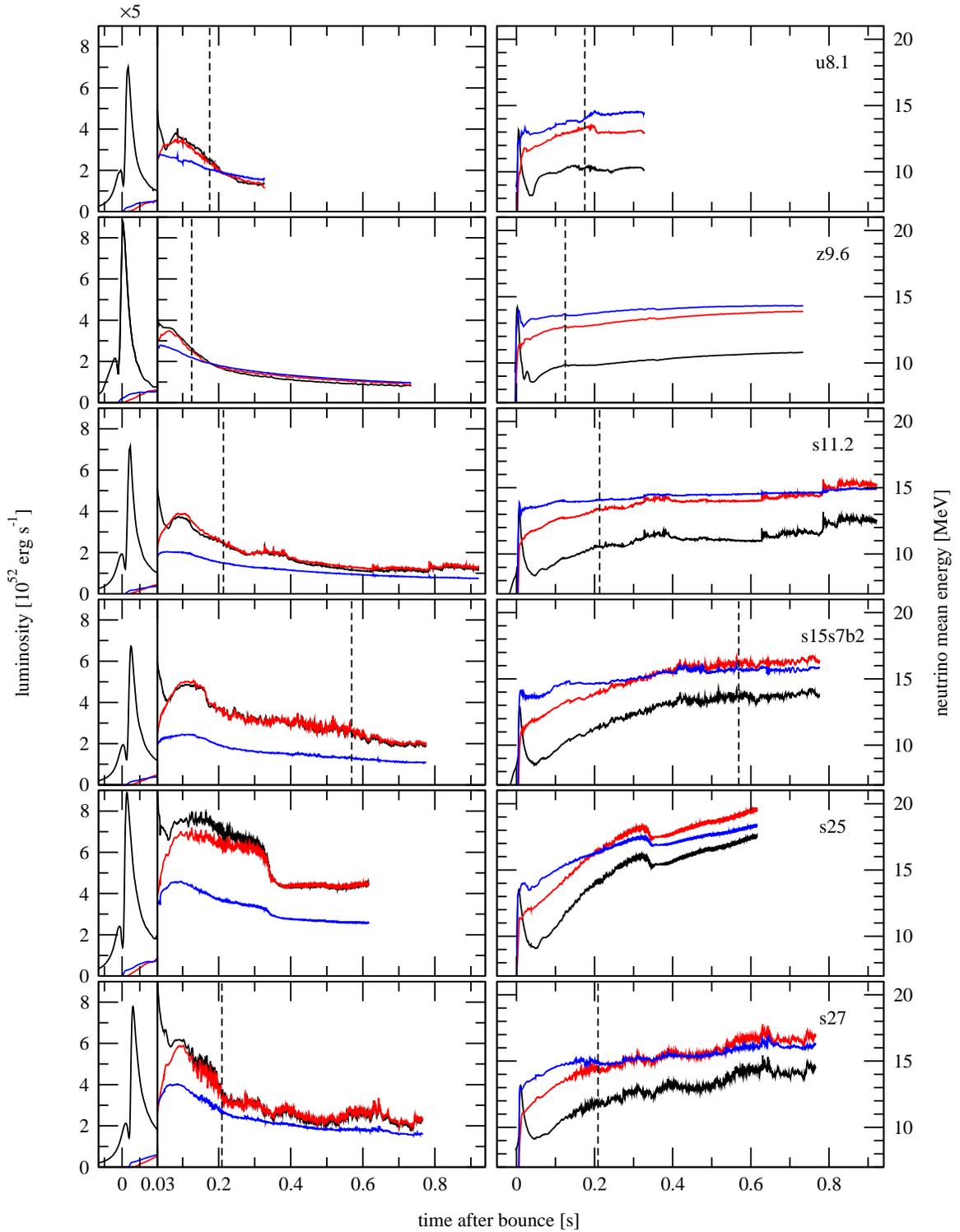}
 \caption{Total energy loss rates (``luminosities`'), $L_\mathrm{tot}$
   (left column), and mean energies, $\langle E \rangle$ (right
   column), of the emitted neutrinos for models u8.1, z9.6, s11.2,
   s15s7b2, s25, and s27 (from top to bottom). Black, red, and blue
   curves are used for electron neutrinos, electron antineutrinos, and
   $\mu/\tau$ neutrinos, respectively. Note that a different scale is
     used prior to a post-bounce time of $30 \, \mathrm{ms}$ in order
     to fit the neutrino shock-breakout burst into the same plot as
     the signal from the accretion phase: During the burst phase, the
     luminosities have been scaled down by a factor of 5, i.e.\ the
     reader should understand that the actual luminosity is higher by that
     factor. A dashed vertical line marks the onset of the explosion
     (defined as the time when the average shock radius reaches
     $400 \, \mathrm{km}$) in exploding models.
   \label{fig:overview}}
\end{figure*}

\begin{figure}
\plotone{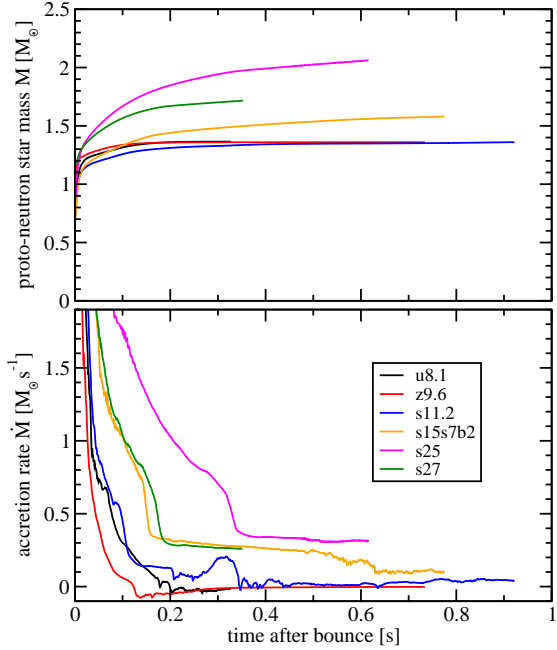}
\caption{ Baryonic proto-neutron star mass $M$ (top) and mass
  accretion rate $\dot{M}$ (bottom, measured at a radius of $500 \,
  \mathrm{km}$) for models u8.1, z9.6, s11.2, s15s7b2, s25, and s27.
  The bumps in $\dot{M}$ around $300\,\mathrm{ms}$ in the case of the
  $11.2\,M_\odot$ model are a consequence of a temporary expansion of
  the shock beyond $500\,\mathrm{km}$.
\label{fig:masses}
}
\end{figure}

\begin{figure}
\plotone{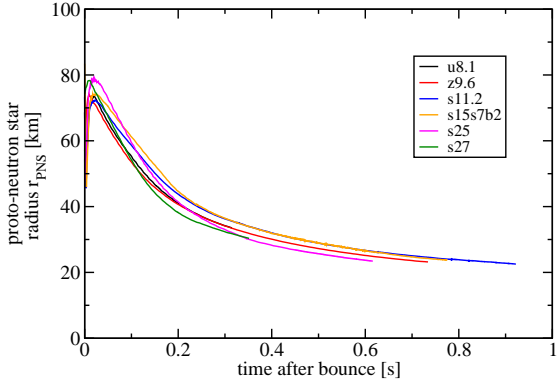}
\caption{ Evolution of the circumferential radius $r_\mathrm{PNS}$
of the proto-neutron star (defined by a fiducial density
of $10^{11} \, \mathrm{g}\, \mathrm{cm}^{-3}$) for models
u8.1, z9.6, s11.2, s15s7b2, s25, and s27.
\label{fig:radii}
}
\end{figure}

\section{Overview of Neutrino Luminosities and Mean Energies}
\label{sec:overview}
The emission of neutrinos from the supernova core is regulated by
physical processes with very different time-scales. The diffusion of
neutrinos from the interior of the hot proto-neutron star to the
neutrinosphere provides a steady, slowly-varying source for neutrinos
of all flavors (and is the dominant source for heavy flavor
neutrinos). This diffusive component originates from the spherically
stratified and roughly isentropic accretion mantle of the
proto-neutron star with densities $\gtrsim 10^{13} \, \mathrm{g}
\, \mathrm{cm}^{-3} $, and is almost isotropic in the absence of
rotation. Electron neutrinos and antineutrinos are also copiously
emitted from the ```cooling region'' outside
the neutrinosphere at optical depth $\lesssim
1$, where fresh material is continually resupplied as long as
accretion onto the proto-neutron star continues. The total accretion
luminosity $L_\mathrm{acc}$ is roughly given by the mass accretion
rate $\dot{M}$ and the gravitational potential at the neutron star
surface \citep{burrows_88,fischer_09},
\begin{equation}
L_\mathrm{acc} \approx \frac{G M \dot{M}}{r_\mathrm{PNS}},
\end{equation}
where $M$ and $r_\mathrm{PNS}$ are the proto-neutron star mass and radius.
$L_\mathrm{acc}$ can vary on much shorter time-scales (milliseconds to tens of
milliseconds) than the diffusive luminosity provided that $\dot{M}$
changes rapidly. If the accretion flow is asymmetric due to convection
or the SASI, the emission from the cooling region can also become
strongly anisotropic. Generally, the emission anisotropies
will be short-lived with a typical time-scale identical to that
of the underlying hydrodynamic instability, i.e.\ a few tens
of milliseconds or less during the accretion phase.

It is expedient to separate the discussion of the neutrino emission on
different temporal and spatial scales. The secular variation of the
angle-integrated ``monopole'' component of the neutrino radiation field
can be analyzed largely without taking the action of convection and
the SASI into account. To this end, we compute the angle-integrated
neutrino energy flux (``total luminosity'' $L_{\mathrm{tot},i}$) of
neutrino flavor $i$,
\begin{equation}
\label{eq:total_luminosity}
L_{\mathrm{tot},i}
=
\int \alpha^2 \phi^4 F_\mathrm{eul} (r=400 \mathrm{km}) r^2\,\ud \Omega,
\end{equation}
and the angle-averaged mean energy $\langle E_i \rangle$,
\begin{equation}
\langle E_i \rangle
=
\frac{\int \alpha^2 \phi^4 F_\mathrm{eul} (r=400 \mathrm{km}) \,\ud \Omega}
{\int \alpha \phi^4 \mathcal{F}_\mathrm{eul} (r=400 \mathrm{km}) \,\ud \Omega},
\end{equation}
for our six 2D models. Here $F_\mathrm{eul}$ and
$\mathcal{F}_\mathrm{eul}$ are the lab-frame neutrino energy flux and
number flux, respectively, and $\alpha$ and $\phi$ are the lapse
function and conformal factor in the CFC metric.  The neutrino
luminosities and mean energies are extracted at a radius of $400 \,
\mathrm{km}$, where they have essentially reached their asymptotic
values. The total neutrino fluxes $L_{\mathrm{tot},i}$ and the
corresponding mean energies $\langle E_i\rangle$ are shown as
functions of time in Figure~\ref{fig:overview}.  { We also show the
  temporal evolution of important parameters of the proto-neutron star
  and the accretion flow, which determine the neutrino luminosities and
  mean energies in Figures~\ref{fig:masses} and
  \ref{fig:radii}. Figure~\ref{fig:masses} presents the time evolution
  of the baryonic mass of the proto-neutron star (defined as the total
  amount of material at densities exceeding $10^{11} \, \mathrm{g} \,
  \mathrm{cm}^{-3}$) and the (baryonic) mass accretion rate $\dot{M}$
  (measured at a radius of $400\, \mathrm{km}$). The contraction of
  the proto-neutron star radius $r_\mathrm{PNS}$ is shown in
  Figure~\ref{fig:radii}.  }

\subsection{Neutrino Burst and Early Accretion Phase}
The first prominent feature in Figure~\ref{fig:overview} is the
well-known electron neutrino or neutronization burst that occurs when
the post-shock matter becomes optically thin shortly after bounce as
the newly-formed shock propagates outwards (```shock breakout''). As
the post-shock region is far away from neutrino-less beta-equilibrium,
electron captures on protons quickly produce a large number of
electron neutrinos at this stage.

The neutronization burst signal in $\nu_e$ is rather homogeneous
across the different progenitors (in agreement with
\citealt{mayle_87,liebendoerfer_03,kachelriess_05}) with a maximum
luminosity of $(3.3 \ldots 4.4) \times 10^{53} \, \mathrm{erg}
\, \mathrm{s}^{-1}$, an interval bracketed by models s15s7b2 and s25 on
the lower and upper end. There is even less spread in the peak mean
energy of electron neutrinos $\langle E_{\nu_e}\rangle$ reached during
the burst, which ranges from $12.9 \, \mathrm{MeV}$ (s15s7b2) to $14.2
\, \mathrm{MeV}$ (z9.6).

\begin{figure}
\plotone{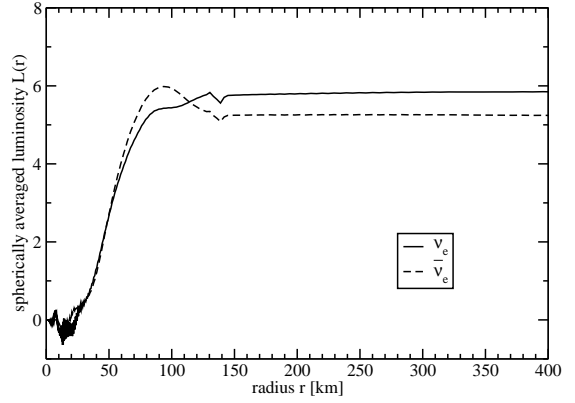}
\caption{Radial profiles of $L_{\nu_e}$ and $L_{\bar{\nu}_e}$ during
  the early accretion phase ($70 \, \mathrm{ms}$ after bounce) for
  model s27. The hierarchy of the luminosities at infinity $L_{\nu_e}
  > L_{\bar{\nu}_e}$ is inverted compared to the luminosities
  just outside the proto-neutron star ($r \approx 90 \, \mathrm{km}$),
  because emission dominates over absorption for electron
  neutrinos in the post-shock region, while the electron antineutrino
  luminosity is reduced considerably between $90 \, \mathrm{km}$ and 
  $140 \, \mathrm{km}$ due to net absorption.
The kink at $\sim 140 \, \mathrm{km}$ coincides with the shock position
and is an artifact of the transformation from the comoving frame
to the lab frame.
\label{fig:profile_s27_70ms}
}
\end{figure}

As the $\nu_e$ burst subsides, the flux of $\bar{\nu}_e$ and
$\nu_{\mu/\tau}$ starts to rise. While the emission of $\bar{\nu}_e$
is initially suppressed due to  a relatively high electron
  fraction around the neutrinosphere and in the cooling region above
it (which implies strong electron degeneracy), leading to a
delayed rise of $\bar{\nu}_e$'s compared to the $\nu_{\mu/\tau}$'s,
the $\bar{\nu}_e$'s eventually reach a flux similar to the
$\nu_e$'s. The excess of electron (anti)\-neutrino emission compared to
the heavy flavor neutrinos is due the contribution from accretion and
is hence progenitor-dependent; for the low-mass progenitors u8.1 and
z9.6, it is never very pronounced, whereas the electron neutrino and
antineutrino luminosities can be larger almost by a factor of two for
more massive progenitors.

\begin{figure}
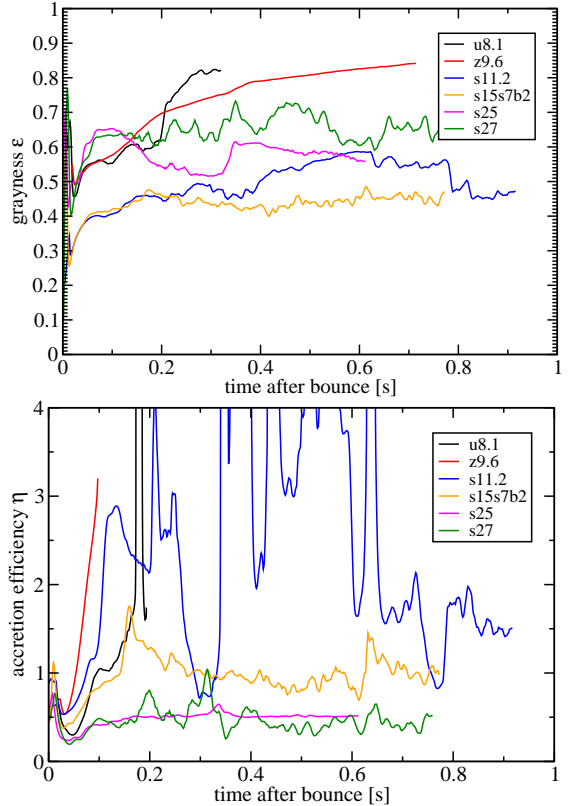

\plotone{f5a.eps}
\plotone{f5b.eps}
\caption{Grayness parameter $\epsilon$ for the emission of $\mu$ and
  $\tau$ neutrinos (top panel) and accretion efficiency $\eta$ (bottom
  panel) for all models. Note that the accretion efficiency is
  computed using the mass accretion rate $\dot{M}$ at $r=500
  \, \mathrm{km}$.
\label{fig:grayness}
}
\end{figure}

\begin{figure}
\plotone{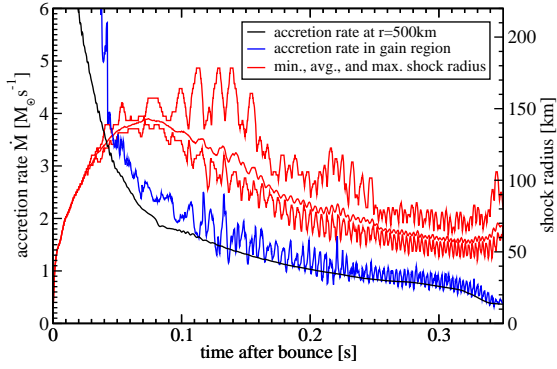}
\caption{Accretion rate $\dot{M}$ ahead of the shock at $r=500
  \, \mathrm{km}$ (black) and between the gain radius and average shock radius
  at $r=(r_\mathrm{gain}+r_\mathrm{sh})/2$ (blue) for model s25. The minimum,
  average and maximum shock radii are also shown (red). The (small)
  periodic expansion and retraction of the average radius is reflected
  in a relatively strong quasi-periodic modulation of the accretion
  rate in the gain region. As a result of the fluctuating
  supply of material into the cooling region, there is a small
  periodic variation in the total neutrino luminosity
(see fifth row in Figure~\ref{fig:overview}). Only the
  first $\sim 350 \, \mathrm{ms}$ after bounce are shown to make
  the fluctuations more clearly visible. Compare
  the fifth row in Figure~\ref{fig:overview} for the resulting
  modulation of the electron neutrino and antineutrino luminosities.
\label{fig:accretion_s25}
}
\end{figure}

\begin{figure}
\plotone{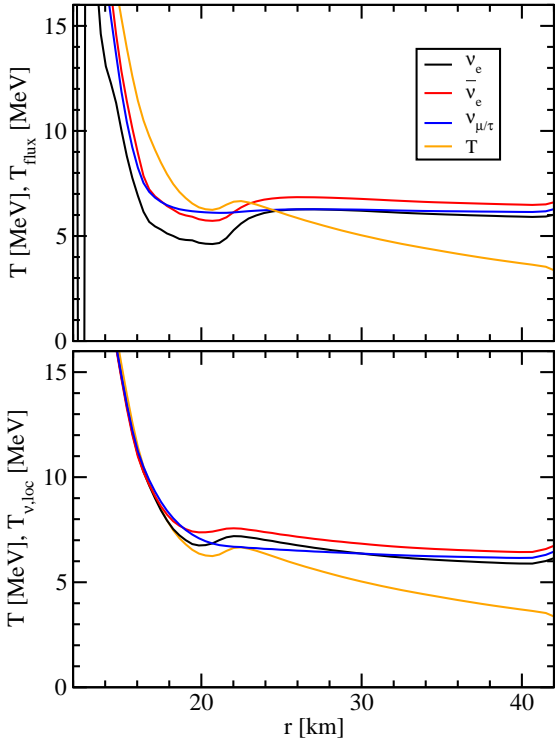}
\caption{Radial profiles of the flux temperature $T_\mathrm{flux}$
  (top panel, Equation~\ref{eq:tflux}) and the local temperature
  $T_{\nu,\mathrm{loc}}$ (bottom panel, Equation~\ref{eq:teff}) for the different
  neutrino species compared to the matter temperature $T$ at a
  post-bounce time of $591 \, \mathrm{ms}$ for model s25. $T_{\nu,\mathrm{eff}}$ follows the
  matter temperature closely deep in the optically thick regime, but
  starts to deviate from $T$ well before its asymptotic value is
  reached. $T_\mathrm{flux}$ is generally lower than $T$ at high
  optical depths. $T_\mathrm{flux}$ approaches or crosses $T$ where
  energy-exchanging reactions freeze out and reaches it asymptotic
  value close to that point.
\label{fig:crossing}
}
\end{figure}

\begin{figure}
\plotone{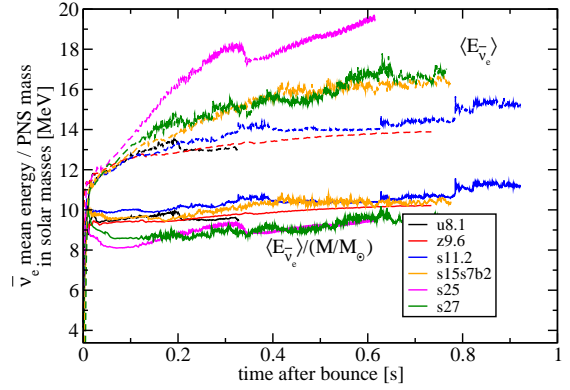}
\caption{Time evolution of the electron antineutrino mean energy
  $\langle E_{\bar{\nu}_e} \rangle / (M / M_\odot)$ rescaled by the
  proto-neutron star mass $M$ for the different progenitors. The mean
  energies $\langle E_{\bar{\nu}_e} \rangle$ themselves are shown as
  dashed lines along with the rescaled mean energies $\langle
  E_{\bar{\nu}_e} \rangle / (M / M_\odot)$ (solid lines). When
  comparing different progenitors the reader should note that
  models s11.2 and s15s7b2 have been computed with an angular
  resolution of $2.8^\circ$ instead of $1.4^\circ$ (u8.1, z9.6, s25,
  s27), and that two different EoS's have been used
  (u8.1, s11.2, s15s7b2: LS180; z9.6, s25, s27: LS 220).
\label{fig:mean_energy_scaled}
}
\end{figure}

\begin{figure}
\plotone{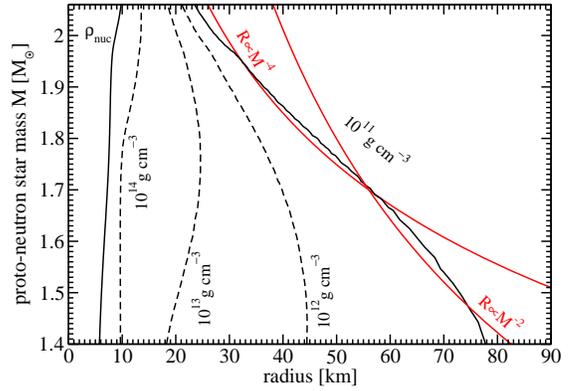}
\caption{ Mass-radius relation for different density isosurfaces as
  observed during the evolution of model s25. The black curves are
  defined by the baryonic mass $M$ of the \emph{entire} proto-neutron
  star (i.e.\ the mass contained in the density isosurface for
$10^{11}\, \mathrm{g}\, \mathrm{cm}^{-3}$) and the radius corresponding to different densities (nuclear
  saturation density $\rho_\mathrm{nuc}$,
  $10^{14}\,\mathrm{g}\,\mathrm{cm}^{-3}$,
  $10^{13}\,\mathrm{g}\,\mathrm{cm}^{-3}$,
  $10^{12}\,\mathrm{g}\,\mathrm{cm}^{-3}$, and
  $10^{11}\,\mathrm{g}\,\mathrm{cm}^{-3}$). The curves for nuclear
  saturation density and the density of
  $10^{11}\,\mathrm{g}\,\mathrm{cm}^{-3}$ (which we use to define the
  proto-neutron star radius in this paper) are shown as solid lines;
  curves for other densities are shown as dashed lines. Curves for
  power laws $R\propto M^{-4}$ and $R\propto{M}^{-2}$ are shown in red
  for comparison. It should be noted that the high-density core
  retains a relatively constant radius as the proto-neutron star
  accretes. The outer layers contract strongly with increasing $M$.
\label{fig:mass_radius}}
\end{figure}

Interestingly, the different progenitors show a different luminosity
hierarchy of electron neutrinos and antineutrinos during the early
accretion phase with an early crossing around $\sim 60 \, \mathrm{ms}$
after bounce in the case of models s11.2 and s15s7b2 where the
$\bar{\nu}_e$ luminosity slightly exceeds the $\nu_e$ luminosity (a
phenomenon observed even more strongly in the $12 M_\odot$ model
simulated by \citealt{bruenn_13}). This is a subtle effect that cannot
be connected to the surface properties of the proto-neutron
star. During this early phase, neutrinos are emitted from a very
extended region, and for electron neutrinos, emission still dominates
over absorption all the way to the shock due to the deleptonization of
the infalling matter. By contrast, a considerable fraction of the
electron antineutrinos is absorbed in the region where net heating
develops later on. The different balance between emission and
absorption in the post-shock region results in a considerable
re-adjustment for the electron flavor luminosities from the
proto-neutron star surface and deeper layers of the cooling region, as
we show in Figure~\ref{fig:profile_s27_70ms} for model s27, which does
not show the early crossing of $\nu_e$ and $\bar{\nu}_e$
luminosities. Moreover, proto-neutron star convection affects electron
neutrino and antineutrino luminosities differently at early stages,
generally tilting the balance in favor of electron neutrinos by
transporting lepton number from deeper layers into the neutrinosphere
region \citep{buras_06_a}.  The hierarchy of electron neutrino and
antineutrino luminosities during the early phase thus seems almost
accidental, and is certainly very sensitive to small variations in the
models (such as the extent of the region affected by proto-neutron
star convection and the exact temporal decay behavior of the mass
accretion rate in the early post-bounce phase).

\subsection{Later Accretion Phase -- Luminosities}
During the later accretion phase, the electron neutrino and
antineutrino luminosities are largely regulated by the mass accretion
rate $\dot{M}$ \citep{fischer_09}. For high $\dot{M}$, the total
electron flavor luminosity is of the order of the accretion luminosity
$L_{\nu_e}+L_{\bar{\nu}_e} \sim G M \dot{M}/ r_\mathrm{PNS}$ ($M$ and
$r_\mathrm{PNS}$ being the mass and radius of the proto-neutron star,
respectively), though it may deviate by a few tens of percent.

As the accretion rate drops, the relative contribution of the electron
neutrino and antineutrino flux from diffusion out of the deeper PNS
core becomes more appreciable. Rapid changes in $\dot{M}$ due to the
infall of composition interfaces in the progenitor
result in pronounced steps in the
luminosity of $\nu_e$ and $\bar{\nu}_e$ (see
Figure~\ref{fig:overview}). The heavy flavor neutrinos show a
steady decline and are not sensitive to sudden variations in
$\dot{M}$.

Although the neutrino luminosities during the accretion phase
are related to the proto-neutron star parameters, quantitative
deductions from the luminosities appear difficult. For the
heavy flavor neutrinos, one can assume a simple gray body ansatz
in terms of the neutron star surface temperature $T_\nu$
and radius $r_\mathrm{PNS}^2$,
\begin{equation}
L_{\nu_{\mu/\tau}}=4 \pi \epsilon \sigma_\mathrm{fermi} r_\mathrm{PNS}^2 T_{\nu}^4.
\end{equation}
Here,
$\sigma_\mathrm{fermi}=
4.50 \times 10^{35} \, \mathrm{erg}\, \mathrm{MeV}^{-4} \mathrm{s}^{-1} \mathrm{cm} ^{-2}$
is the radiation
constant for neutrinos as left- or right-handed fermions
of zero degeneracy,
$T_{\nu}$ is an effective surface temperature approximately given by
$\langle E_{\bar{\nu}_e} \rangle / 3.15$, and the grayness factor $\epsilon$
accounts for deviations from black body emission. Figure~\ref{fig:grayness} shows a
considerable spread in $\epsilon$ across the progenitors as well as
strong non-monotonic time variations. We obtain values in the range
$\epsilon = 0.4 \ldots 0.85$, which is compatible with the findings of
\citet{huedepohl_10} for an electron-capture supernova during the
first second.

Similarly, the accretion rate $\dot {M}$ can only be inferred very
tentatively from the luminosities. The accretion luminosity
$L_\mathrm{acc}$ is already difficult to separate from the diffusive
(gray body) contribution to begin with.  As long
as it is sufficiently
high, the excess of the electron (anti)\-neutrino luminosity 
over the (anti)\-neutrino luminosity of a single heavy
flavor may be taken as a
proxy for $L_\mathrm{acc}$, which may then be related
to the accretion rate $\dot{M}$ by introducing
an efficiency parameter $\eta$,
\begin{equation}
L_\mathrm{acc}\approx L_{\nu_e}+L_{\bar{\nu}_e}-2 L_{\nu_{\mu/\tau}} = \eta \frac{G M \dot{M}}{ r_\mathrm{PNS}}.
\end{equation}
The bottom panel of Figure~\ref{fig:grayness} shows that for the less
massive progenitors (u8.1, z9.6, s11.2) the subtraction of a diffusive
component works up to shock revival at best, where $\eta \approx
1$. Afterwards, $\eta$ increases steeply as the accretion rate
$\dot{M}$ plummets. For models with an extended accretion phase
continuing beyond the onset of the explosion (s15s7b2, s25, s27), we
find $\eta \approx 0.5 \ldots 1$ most of the time even after shock
revival.

It is noteworthy that even the (directionally averaged) luminosity shows
marked temporal fluctuations in models with strong activity of the
SASI and/or convection, such as s15s7b2, s25, and s27. These
fluctuations are a consequence of oscillations in the angle-averaged
shock radius, resulting in a periodic increase and decrease of the
rate of mass accretion $\dot{M}_\mathrm{cool}$ into the cooling region as
shown in Figure~\ref{fig:accretion_s25}. However, in the
observable signal for a fixed viewing angle these variations
in the angle-integrated luminosity are dwarfed by spatio-temporal
variations in the neutrino emission, which are discussed
at length in Section~\ref{sec:variations}.

\subsection{Later Accretion Phase -- Mean Energies}
\label{sec:crossing}

The directionally averaged neutrino mean energies show an almost
monotonic rise. The infall of composition interfaces, however, can
lead to short phases of stagnation or even drops in the mean
energies, but the mean energies are much less affected by changes in
the accretion rate than the luminosities.

In all but the two least massive progenitors (u8.1 and z9.6), we
eventually observe a crossing of the mean energies
of electron antineutrinos and heavy flavor neutrinos, and hence
a violation of the canonical hierarchy of mean energies
$\langle E_{\nu_{\mu/\tau}}\rangle > \langle E_{\bar{\nu}_{e}}\rangle
> \langle E_{\nu_e}\rangle$. This feature has been found
by many modern 1D and 2D neutrino-hydrodynamics simulations
\citep{liebendoerfer_04,marek_08,marek_09} that include some or all of the relevant thermal equilibration
processes for heavy flavor neutrinos (neutrino-electron
scattering, nucleon bremsstrahlung, neutrino pair conversion,
energy-exchanges in neutrino-nucleon scattering reactions).

These processes are one important cause for the crossing as they push
the effective heavy flavor temperature $T_{\nu_{\mu/\tau},\mathrm{eff}} \approx
\langle E_{\nu_{\mu/\tau}}\rangle / 3.15$ down close to the matter
temperature in the roughly isothermal atmosphere of the proto-neutron
star (where most of the electron neutrinos and antineutrinos are
produced prior to the explosion), a process discussed at length in
\citet{raffelt_01,keil_03}. However, this does not yet account for the
crossing of the mean energies, which hinges on two other factors that
are illustrated in Figure~\ref{fig:crossing}.
Figure~\ref{fig:crossing} shows the evolution of the mean \emph{flux}
temperature $T_\mathrm{flux}$ and the local spectral
temperature $T_{\nu,\mathrm{loc}}$ of the different neutrino species as they
propagate out of the proto-neutron star atmosphere compared to the
matter temperature at a time well after the crossing for model s25.  The flux
temperature $T_\mathrm{flux}$ for species $i$ is defined in terms of
the 1st angular moment $H$ of the energy-dependent neutrino intensity in the
comoving frame as
\begin{equation}
\label{eq:tflux}
T_\mathrm{flux}
=
\frac{1}{3.15}
\frac{\int_0^\infty H \, \ud E_\nu}{\int_0^\infty E_\nu^{-1} H \, \ud E_\nu},
\end{equation}
i.e.\ the temperature $T_\mathrm{flux}$ is obtained from ratio of the
neutrino energy and number flux assuming that the spectrum is a
Fermi-Dirac spectrum with vanishing chemical potential. The local
neutrino temperature $T_{\nu,\mathrm{loc}}$ is defined similarly in
terms of the 0th moment $J$ as
\begin{equation}
\label{eq:teff}
T_{\nu,\mathrm{loc}}
=
\frac{1}{3.15}
\frac{\int_0^\infty J \, \ud E_\nu}{\int_0^\infty E_\nu^{-1} J \, \ud E_\nu}.
\end{equation}
$T_\mathrm{flux}$ reflects the thermodynamic conditions in
the decoupling region, and with a negative temperature gradient in the
outer neutron star atmosphere one could naturally expect the standard
hierarchy, since the heavy flavor neutrinos decouple at smaller radii
than electron antineutrinos.

However, Figure~\ref{fig:crossing} demonstrates a loophole in this
argument. As already pointed out by \citet{liebendoerfer_04}, a
\emph{temperature inversion} can occur in the cooling region, where
$\nu_e$ and $\bar{\nu}_e$ are produced by charged-current
processes. The flux temperature of electron antineutrinos can therefore
be higher than that of the heavy flavor neutrinos. The asymptotic
value of $T_{\bar{\nu}_e,\mathrm{flux}}$ roughly corresponds to the
local temperature maximum at $\sim 22 \, \mathrm{km}$, whereas the flux
temperature of $\nu_{\mu/\tau}$'s reaches its asymptotic value around
the local temperature minimum at $\sim 20 \, \mathrm{km}$, which is
about $0.5 \, \mathrm{MeV}$ lower.  The temperature inversion can
easily be accounted for; it is a result of the competition between
neutrino cooling and adiabatic compression in the accretion
layer. Assuming spherical symmetry, stationarity, and Newtonian
gravity, and neglecting self-gravity we can write the energy equation
for the fluid in terms of the radial velocity $v_r$, the specific
enthalpy $h$, the gravitational potential $\Phi$, and the source term
$\dot{q}$ due to neutrino heating/cooling as
\begin{equation}
\frac{\ud (h +\Phi)}{\ud t}=
v_r \frac{\pd (h +\Phi)}{\pd r}= \frac{\dot{q}}{\rho}.
\end{equation}
With baryon-dominated gas in the cooling region, the specific enthalpy is
approximately $5/2 k_b T/m_n$ ($m_n$ being the nucleon mass) so that the
gradient of $h$ immediately reflects the temperature gradient, and the
competition between cooling and adiabatic compression becomes obvious (note
that both $\dot{q}$ and $v_r$ are negative in the cooling region):
\begin{equation}
\label{eq:temperature_gradient}
\frac{5 k_b}{2m_n}\frac{\pd T}{\pd r}\sim \frac{\dot{q}}{v_r}-\frac{\pd \Phi}{\pd r}.
\end{equation}
As the density gradient in the atmosphere steepens, the accreted
matter has to radiate away its gravitational binding energy while traversing
an increasingly narrower cooling region. The balance is then tilted in favor
of the term $\frac{\dot{q}}{v_r}$ in Equation~(\ref{eq:temperature_gradient}),
and a positive temperature gradient in the outer proto-neutron star atmosphere arises.

Moreover, electron (anti)\-neutrinos of different energies decouple at
different radii because the absorption and scattering cross sections
scale with the square of the neutrino energy, $E_\nu^2$. The emerging
energy-dependent luminosity $\ud L/\ud E_\nu$ is essentially
determined by the equilibrium intensity $\mathcal{I}_\mathrm{eq}$ at
an optical depth of $\sim 2/3$ and by the effective emitting surface,
\begin{equation}
\frac{\ud L}{\ud E_\nu}
\propto
r(\tau=2/3)^2 \times \mathcal{I}_\mathrm{eq}(\tau=3/2,E_\nu).
\end{equation}
Because of the relatively strong variation of
$\mathcal{I}_\mathrm{eq}$ with temperature
$\mathcal{I}_\mathrm{eq} \propto E_\nu^3 [1+\exp (E_\nu/k_b T- \eta_\mathrm{eq})]$, the emission of neutrinos
with energies near the spectral maximum in the region around the local
temperature maximum is enhanced compared to that of low-energy
neutrinos that decouple further inside and thus at lower temperatures. As a
result, the mean energy of the emerging non-thermal spectrum can be
even higher than one would assume based on the temperature in the
decoupling region (cf.\ Figure~\ref{fig:crossing}, lower panel). 

Neutrinos in the high-energy tail, on the other hand, are emitted from
regions where there is again an appreciable negative temperature
gradient, and the spectrum therefore declines faster than a thermal
spectrum for neutrinos well above the average energy. If the
temperature inversion in the accretion layer is strong, this ``pinching''
\citep{janka_89a,janka_89b,raffelt_01,keil_03,tamborra_12} is no longer effective enough to keep the
electron antineutrino mean energy below that of the heavy flavor
neutrinos during the late accretion phase. The root mean square energy,
$\langle E_\nu^2\rangle^{1/2}$, however, is more affected by the
high-energy tail and always shows the canonical hierarchy $\langle
E_{\nu_e}^2\rangle^{1/2} < \langle E_{\bar{\nu}_e}^2\rangle^{1/2}<\langle E_{\nu_{\mu/\tau}}^2\rangle^{1/2}$
(cf.\ also \citealt{marek_09}).

We should also mention that the precise instant of the crossing is
probably affected by weak magnetism corrections \citep{horowitz_02},
which are different for $\nu_{\mu/\tau}$ and $\bar{\nu}_{\mu/\tau}$.
Weak magnetism lowers the opacity for antineutrinos and leads to
somewhat harder spectra (by about $1 \, \mathrm{MeV}$;
\citealt{liebendoerfer_03,bruenn_13}), and this effect would also
occur in $\bar{\nu}_\mu$ and $\bar{\nu}_\tau$. The \textsc{Vertex}
transport code does not distinguish heavy flavor neutrinos and
antineutrinos in its current implementation, so that we obtain mean
energies for $\bar{\nu}_{\mu/\tau}$ that are slightly too low.  However,
simulations that treat heavy flavor neutrinos and antineutrinos
separately still show the crossing despite slightly higher heavy
flavor antineutrino mean energies, albeit at slightly different times
for $\nu_{\mu/\tau}$ and $\bar{\nu}_{\mu/\tau}$.

\subsection{Relation of Neutrino Mean Energies and Proto-Neutron Star Parameters}
The gradual rise of the neutrino mean energies is clearly a consequence
of the contraction of the proto-neutron star and its rising surface
temperature. Surprisingly, little attention has been paid to the
dominant parameters regulating the neutrinospheric temperature during
the accretion phase so far. Figure~\ref{fig:mean_energy_scaled} shows
that for each model, the electron antineutrino mean energy scales
remarkably well with the proto-neutron star mass $M$ (defined by
the baryonic mass contained in regions with densities exceeding
$10^{11} \, \mathrm{g} \, \mathrm{cm}^{-3}$),
\begin{equation}
\label{eq:mass_energy}
\langle E_{\bar{\nu}_e}(t) \rangle \propto M(t).
\end{equation}
The transient stagnation or decrease of the mean energy associated
with the infall of composition interfaces causes slight deviations from
the relation. There is also a spread of $\lesssim 20\%$ between
different progenitor models such that more massive progenitors (s25 and
s27) tend to have a somewhat lower ratio $\langle E_{\bar{\nu}_e}(t)
\rangle / M(t)$. To first order, Equation~(\ref{eq:mass_energy}) still accounts
reasonably well for the large variation of the neutrino mean energies
($11 \, \mathrm{MeV}$ to $20 \, \mathrm{MeV}$ for $\bar{\nu}_e$) through
the post-bounce phase for different progenitors.

How can this scaling be accounted for, and how can it be reconciled
with the fact that the nuclear equation of state also affects the
neutrino mean energies considerably
\citep{sumiyoshi_05,marek_08,marek_09,oconnor_13}? In order to answer
this question, we construct a simple model for the isothermal
atmosphere of the proto-neutron star whose properties determine the
emerging electron neutrino and antineutrino spectrum.

We first note that the transition from a steep negative temperature gradient
in the roughly isentropic accretion mantle of the proto-neutron star
to an almost flat temperature gradient in the cooling region (Figure~\ref{fig:crossing}) roughly
coincides with the neutrinosphere for $\bar{\nu}_e$ (simply
because efficient cooling can only occur at low optical depth $\tau$).
At the neutrinosphere radius $R_\nu$, we have
\begin{equation}
\label{eq:atmosphere_boundary}
\tau=\int_{R_\mathrm{\nu}}^\infty \kappa_\mathrm{eff} \,\ud r \approx 1,
\end{equation}
where $\kappa_\mathrm{eff}$ is the effective energy-averaged opacity for electron
antineutrinos. Since the dominant absorption and scattering opacity
scales with the density $\rho$ and the neutrino energy $E_\nu$ as
$\rho E_\nu^2$, this implies
\begin{equation}
\label{eq:atmosphere_boundary2}
C \int_{R_\mathrm{\nu}}^\infty \rho T_\nu^2 \,\ud r \approx 1,
\end{equation}
where $T_\nu$ is the temperature at the average neutrinosphere (and
the effective temperature of the emitted electron antineutrinos), and
$C$ is a normalization constant.  The density $\rho(r)$ in this
integral is in turn determined by the density $\rho_\nu$ at the base
of the { roughly isothermal, baryon-dominated} atmosphere { (cooling region and neutrinospheric region)}, the temperature $T_\nu$, and the local
gravitational acceleration $GM/R_\nu^2$ at $R_\nu$ 
(cp.\ \citealp{mihalas_84}, Chapter~2),
\begin{equation}
\label{eq:density_atmosphere}
\rho \approx \rho_\nu \exp \left(-\frac{G M \left(r- R_\nu\right)
  m_n}{R_\nu^2 k T_\nu} \right).
\end{equation}
Using these relations,\footnote{Note that the integral in Equations~(\ref{eq:atmosphere_boundary}), (\ref{eq:atmosphere_boundary2})
is dominated by the contributions from regions close to the neutrinosphere
where Equation~(\ref{eq:density_atmosphere}) holds.} we can determine that the effective
optical depth $\tau_\nu$ for electron antineutrinos
at the base of the atmosphere scales as
\begin{equation}
\tau_\nu \propto \frac{\rho_\nu R_\nu^2 T_\nu^3}{G M}.
\end{equation}
$\rho_\nu$ can be eliminated from this relation if we bear in mind
that for $r<R_\nu$, the stratification in the thick accretion mantle
is roughly adiabatic because it is well-mixed by proto-neutron star
convection.  For densities sufficiently below nuclear saturation
density and down to $\rho_\nu$, { where the nearly isentropic mantle
  smoothly joins the isothermal atmosphere}, the plasma can be
described as a mixture of relativistic and non-relativistic ideal gas
components, and the adiabaticity constraint therefore leads to a
power-law for the temperature,
\begin{equation}
\label{eq:isentropic}
T = K(s) \rho^{\gamma-1},
\end{equation}
where $\gamma \approx 3/2$ is the effective adiabatic index and $K(s)$ is a
constant depending solely on the entropy $s$. The specific entropy in
the accretion mantle (and hence $K(s)$) does not vary considerably
across progenitors and changes very slowly during the accretion phase
for the following reasons: The entropy profile during the early
post-bounce phase is determined by the shock propagation through the
iron core, which is rather similar for different progenitors. During
the subsequent post-bounce evolution, the matter piled up on
the proto-neutron star surface then cools down to the bulk
entropy of the accretion matter as it settles, and essentially
just extends the accretion mantle with constant entropy. The rather
high mass of the mantle and a thermal relaxation (cooling) time-scale
on the order of seconds also help to keep the entropy
relatively constant.

After eliminating $\rho_\nu$, the condition $\tau_\nu \approx 1$ {
  at the neutrinosphere} translates into a relation between the
neutrinosphere temperature and the surface gravity at the base of the
atmosphere:
\begin{equation}
\label{eq:t_surface_gravity}
T_\nu^5 \propto \frac{G M}{R_\nu^2}.
\end{equation}
The neutrinosphere radius $R_\nu^2$ is essentially determined by the
proto-neutron star mass only through the mass-radius relationship for
neutron stars with a cold inner core ($s\approx 1 \,
\mathrm{k_b}/\mathrm{nucleon}$) of $\approx 0.43 \ldots 0.5 M_\odot$
(depending on the EoS) and an adiabatically stratified mantle with $s
\approx 5 \mathrm{k_b}/\mathrm{nucleon}$.  { $R_\nu$ is a
  relatively steep function of the proto-neutron star mass, in
  contrast to cold neutron stars, where $R \approx \mathrm{const.}$
  over a wide range of masses for nucleonic EoSs (see, e.g., Figure~1
  in \citealt{mueller_12a} for an example with a mantle of relatively
  low entropy).  Indeed, the steepness of the mass-radius relation for
  cold neutron stars partly helps to understand why proto-neutron stars with
  a hot mantle contract strongly with increasing mass, as we shall see
  in the following.

In the core of a proto-neutron star above nuclear saturation
density, repulsive nucleon-nucleon interactions dominate the pressure,
and thermal effects are subdominant. Down to and even
somewhat below nuclear saturation density, the density profile
of the proto-neutron star therefore roughly follows a TOV
solution for a cold neutron star. Somewhat below nuclear saturation
density, the thermal pressure of the baryons becomes the dominant
contribution, and the density profile transitions
into the TOV solution for an adiabatic mantle, with a much more
shallow decline of the density compared to cold neutron stars
at comparable densities.

Since the density profile of cold neutron stars is very steep below
saturation density, the ``fitting radius'' between the core and the
mantle is always close to the radius of a cold neutron star with a
baryonic mass identical to that of the core. The steepness of the
mass-radius relation for cold neutron stars
\citep{lattimer_07,steiner_10} implies that the radius of the inner
core remains roughly constant as the proto-neutron star accretes. This
is illustrated in Figure~\ref{fig:mass_radius}, which shows the
evolution of isosurfaces corresponding to nuclear saturation density
and lower fiducial densities in the neutron star for model s25.

The outer radius of the adiabatic mantle on top of the core, i.e.\
$R_\nu$, not only
depends on the radius of the inner core, but also on the
gravitational acceleration in the mantle: The pressure at the base of
the mantle, $P_\mathrm{core}$, remains fairly constant as the
proto-neutron star accretes; it is roughly given by the pressure of
cold neutron star matter at the fitting density. As the proto-neutron star
grows in mass and the surface gravity increases, only a more and more
tenuous mantle can be supported by $P_\mathrm{core}$, and therefore the mantle
contracts. More quantitatively, we can argue that the extent of the
mantle is regulated by the pressure scale-height $\ell$, which is inversely
proportional to the mass of the supranuclear core $M_\mathrm{core}$,
$\ell \propto P_\mathrm{core} /(M_\mathrm{core}^{-1})$.
 Since more and more of the proto-neutron star mass is
concentrated in the core as its total baryonic mass increases, and
since the self-gravity of the mantle also plays a role, $R_\nu$
actually declines even faster with $M$ than $M^{-1}$.
The rapid contraction of isosurfaces for densities  of $\sim 10^{11\ldots 12} \, \mathrm{g}\, \mathrm{cm}^{-3}$,
is also shown in Figure~\ref{fig:mass_radius}.

Of course, the precise dependence of $R_\nu$ on $M$ can only be determined by the
solution of the TOV equation either for a pre-defined stratification
or by using the dynamically evolving stratification in supernova
simulations in conjunction with a solution of the stationary neutrino
transport equations (which is both implicitly done as we evolve the
models). Empirically, it turns out that well below the maximum
neutron star mass, the solutions of the TOV equation (which is
implicitly reproduced in our time-dependent simulations) for such a
stratification can be approximated by a power law with index
\begin{equation}
\label{eq:mass_radius}
\frac{\ud \ln R_\nu}{\ud \ln M} \sim -2,
\end{equation}
as shown by Figure~\ref{fig:mass_radius}.
This value for the power-law index suggests precisely the mass-temperature relation we find in our simulations,
\begin{equation}
\label{eq:mass_temperature}
T_\nu \propto M.
\end{equation}
The actual TOV solution is, of course, \emph{not} a power law.
Nonetheless, deviations from this power-law behavior do not change
the mass-temperature relation appreciably for two reasons: First, different
power laws still give relatively similar neutron star radii
simply because the proto-neutron star mass range is limited
(cp.\ Figure~\ref{fig:mass_radius}).
Furthermore, the high power of the temperature
in Equation~(\ref{eq:t_surface_gravity}) implies that even a relative deviation
of $30\%$ between different power laws is reduced to a deviation
of $11\%$ in $T_\nu$.}

Naturally, this mass-temperature relation is modified by slight
progenitor variations, by the varying strength of the accretion effect
described in Section~\ref{sec:crossing}, and by gravitational
redshift. Progenitor variations in the mass-temperature relation can
be traced to slightly different entropies in the proto-neutron star
convection zone. Higher entropies in more massive progenitors (s25,
s27) lead to somewhat lower neutrino mean energies.

The equation of state is another important factor that regulates the
neutrino mean energies through the compactness of the proto-neutron
star \citep{sumiyoshi_05,marek_08,marek_09,oconnor_13}. This is not
inconsistent with the mass-temperature relation
(\ref{eq:mass_temperature}). Different equations of state will lead to
different mass-radius relations for the core and mantle, and hence
affect the surface gravity at the neutrinosphere. Stiffer equations of
state generally result in a smaller core mass, a more massive and
extended mantle, and hence somewhat smaller neutrinosphere
temperatures. However, this affects only the proportionality constant
in the mass-temperature relation (\ref{eq:mass_temperature}).  {
  Since relativistic effects in the mantle are only of moderate
  magnitude and equation-of-state variations mainly change the radius
  of the core dominated by nucleon interactions, similar power-laws
  are expected for different equations of state as long as the core
  radius varies little with mass as suggested by neutron star radius
  measurements \citep{lattimer_07,steiner_10}.}  Even a different
slope of the mass-radius relationship for proto-neutron stars with a
hot accretion mantle does not strongly affect the relation expressed
by Equation~(\ref{eq:mass_temperature}) because the neutrino
temperature enters as $T_\nu^5$ in the underlying relation
(\ref{eq:t_surface_gravity}). 1D simulations with the EoS of
\citet{shen_98} indeed show a similar scaling relation for the
electron antineutrino mean energy (H\"udepohl~et~al., in preparation).

\subsection{The Explosion Phase}
\label{sec:explosion_spherical_average}
Different from artificial 1D explosion models, we find no clear
fingerprint of shock revival in the angle-averaged neutrino
luminosities and mean energies. Specifically, there is no abrupt drop
in the electron neutrino and antineutrino luminosity associated with
the onset of the explosion. Models where the shock moves out very
rapidly and accretion is quenched rather abruptly (like u8.1 or z9.6)
also have a rather low accretion luminosity to begin with so that a
strong decline of the luminosity cannot be expected. For other
progenitors (s11.2 and s27), the onset of the explosion is associated
with a strong decline of the luminosity, but this is just a reflection of
the infall of the shell interface that triggers shock revival.
Moreover, the decline of the electron neutrino and antineutrino
luminosity can also be rather gradual because much of the shocked material
is still channeled down to the proto-neutron star through accretion
downflows in these cases. In cases where the explosion is not associated
with the infall of a shell interface (model s15s7b2), or where
the accretion rate remains high (e.g. because of a strongly
unipolar explosion geometry), there may even be no trace at all
of shock revival in the total neutrino luminosity.

The angle-averaged mean energies of electron neutrinos and
antineutrinos may show only weak fingerprint of shock revival: For
models u8.1, z9.6, and s11.2, the mean energies stagnate
or even decline slowly over extended periods ($> 100 \, \mathrm{ms}$),
implying that there is almost no further accretion onto
the proto-neutron star. However, there is still a considerable
rise in the mean energies in models s15s7b2 and s27 due
to the high mass accretion rates after shock revival.

In models s11.2 and s27, the explosion phase is characterized
by stronger non-monotonic variations in the total luminosity and
the angle-averaged mean energies than during the accretion phase.
Such variations on intermediate time-scales are also associated
with anisotropic neutrino emission and are discussed
at length in Section~\ref{sec:variations_explosion}

\section{Spatio-temporal Variations in the Neutrino Emission and the Observable Neutrino Signals}
\label{sec:variations}
Nonradial instabilities like convection and the SASI as well
as global asymmetries during the explosion phase lead to anisotropic
neutrino emission. As a result, the observable neutrino luminosities
and mean energies become strongly direction-dependent and deviate
considerably from the directionally averaged luminosities and mean energies
discussed in Section~\ref{sec:overview}. In this Section, we analyze
the time-frequency structure of the observable neutrino signal and
elucidate how it can reveal detailed time-dependent information
about the dynamics in the supernova core during the accretion
and explosion phase.

\subsection{Reconstruction of the Observable Neutrino Signal}
\label{sec:reconstruction}
 Since the ray-by-ray-plus transport does not account for lateral
  neutrino flux components { outside the neutrinosphere}, the information on each ``ray'' 
  for angle $(\theta,\varphi)$ in 3D or latitude $\theta$ in 2D can
  only propagate radially. In reality, an observer outside the
  neutrinosphere would receive flux not just from one angular zone
  (i.e.\ ray), but from the whole radiating surface facing him. The
observable neutrino luminosity along a specific direction
$(\theta,\varphi)$ therefore differs from the asymptotic value of the
``ray luminosity'' $4\pi \alpha^2 \phi^4
F_\mathrm{eul}(r,\theta,\varphi) r^2$ for ($r \rightarrow \infty$) in
the integrand of Equation~(\ref{eq:total_luminosity}). For this
reason, we reconstruct the observable signal from the ray-dependent
neutrino fluxes following the method of \citet{mueller_e_12}, which
assumes a neutrinospheric emission law $\mathcal{I}(\mathbf{n})
\propto 1+3/2 \cos \gamma (\mathbf{n})$ for the neutrino intensity as
a function of the angle $\gamma$ between the radial direction and the
direction vector $\mathbf{n}$ of the emitted neutrinos. The observable
luminosity $L_o (\mathbf{n})$ at infinity along the direction
$\mathbf{n}$ is then given by the integral of
$\mathcal{I}(\mathbf{n})$ over the visible emitting surface (surface
elements $\ud A$),
\begin{equation}
L_o (\mathbf{n})
=
2 \int\limits_\mathrm{vis.surf.} F_\mathrm{eul} (\mathbf{r},\theta) \left(1+\frac{3}{2} \cos \gamma (\mathbf{n}) \right) \cos \gamma (\mathbf{n})  \,\ud A.
\end{equation}
Observable expectation values $\langle E_o^i (\mathbf{n}) \rangle$ of
powers of the neutrino energy $E$ can be obtained in a similar manner
from the first angular moment of the neutrino radiation intensity:
\begin{equation}
\langle E_o^i (\mathbf{n}) \rangle
=
\frac{ \int\limits_\mathrm{vis.surf.} \int\limits_0^\infty E^{i-1} H (E_\nu,\mathbf{r},\theta) \left(1+\frac{3}{2} \cos \gamma \right) \cos \gamma  \, \ud E \, \ud A}
{ \int\limits_\mathrm{vis.surf.} \int_0^\infty E^{-1} H (E_\nu,\mathbf{r},\theta) \left(1+\frac{3}{2} \cos \gamma\right) \cos \gamma  \, \ud E \, \ud A}.
\end{equation}

In order to study the observability of temporal variations in the
neutrino emission, we use $L_o (\mathbf{n})$ and $\langle E_o^i
(\mathbf{n}) \rangle$ to estimate the expected signal in IceCube
\citep{abbasi_11,salathe_12}, which is best suited for detecting fast
time variations among operating detectors thanks to its excellent
temporal resolution and to the prospective high event rate
(cp.\ \citealp{lund_10,lund_12,tamborra_13}).  Since a simplified
detector model is fully sufficient for the purpose of demonstration, we estimate the
excess rate $\mathfrak{R}$ per time bin ($\Delta t = 1.6384
\, \mathrm{ms}$) over the background due to supernova neutrinos
following \citet{halzen_09}:
\begin{equation}
\mathfrak{R}=186 \, \mathrm{bin}^{-1}
\frac{\tilde{L}_o}{10^{52} \, \mathrm{erg} \, \mathrm{s}^{-1}} 
\frac{  \langle \tilde{E}_o^3 \ \rangle /\langle \tilde{E}_o\rangle } {225 \, \mathrm{MeV}^2}
\left(
 \frac{10 \, \mathrm{kpc}}{d}
\right)^2.
\end{equation}
Here, $\tilde{L}_o$, $\tilde{E}_o^3$, and $\tilde{E}_o$ are the electron
antineutrino luminosity and the third and first energy moment of the 
distribution function as measured on Earth, and $d$ is the distance
to the supernova. The background rate $\mathfrak{R}_0$ is taken to be
\begin{equation}
\mathfrak{R}_0=2200 \, \mathrm{bin}^{-1}.
\end{equation}

The luminosity $\tilde{L}_o$ and the energy moments $\langle
\tilde{E}_o^i \rangle$ of electron antineutrinos emitted from the
supernova may be modified by {MSW { resonances} in the outer
shells of the progenitor and by non-linear collective neutrino flavor
conversion (see \citealp{duan_09,duan_10} for a review).  These
effects partly depend on unknown neutrino parameters (mass hierarchy);
and non-linear flavor conversion is not yet completely understood, in
particular in the absence of axial symmetry in the radiation field
\citep{mirizzi_13,raffelt_13b}.  As a full exploration of the possible
scenarios is beyond the scope of this paper, we estimate the IceCube
signal under the optimistic assumption of a normal mass hierarchy and
neglect non-linear flavor conversion (which is probably suppressed
until the later neutrino-driven wind phase, see
\citealt{chakraborty_11a,sarikas_12a,sarikas_12b}).  Furthermore, we
neglect any possible reconversion of neutrinos due to the Earth
effect.  Under these assumptions, the observable electron antineutrino
luminosity and the energy moments relevant for the IceCube signal can
be expressed in terms of the unmodified quantities $L_o$, $\langle E_o
\rangle$ and $\langle E_o^3 \rangle$ (cp. \citealt{kachelriess_05})
\begin{equation}
\tilde{L}_o =
\cos^2 \theta_{12} L_{o,\bar{\nu}_e} + \sin^2 \theta_{12} L_{o,\bar{\nu}_x},
\end{equation}
\begin{equation}
\tilde{E}_o =
\cos^2 \theta_{12} \langle E_{o,\bar{\nu}_e} \rangle +
\sin^2 \theta_{12} \langle E_{o,\bar{\nu}_x} \rangle,
\end{equation}
\begin{equation}
\tilde{E}_o^3 =
\cos^2 \theta_{12} \langle E_{o,\bar{\nu}_e}^3 \rangle +
\sin^2 \theta_{12} \langle E_{o,\bar{\nu}_x}^3 \rangle,
\end{equation}
where $\sin^2 \theta_{12}=0.311$ \citep{pdg}.  While we cannot hope to
discuss the possible impact of flavor conversion on our findings in
its entirety, this simplified approach at least reflects the fact that
detectors will never measure the unoscillated $\bar{\nu}_e$ signal
computed in our simulations.  Using the unoscillated neutrino flux
would result in a systematic overestimation of the signal-to-noise
ratio of features in the time-frequency domain. Our simplified flavor
conversion is intended to roughly represent the most optimistic case
{(maximum amplitude of fluctuations in the observed electron antineutrino flux)} that can possibly be encountered.

The actual detector signal will also be subject to statistical
fluctuations. We therefore compute simulated IceCube signals assuming a
Poisson distribution with an expectation value of
$\mathfrak{R}(t)$ in each bin. Such simulated signals for the four
progenitor models s11.2, s15s7b2, s25, and s27 for observers located
in the north and south polar directions and in the equatorial plane are
shown in Figure~\ref{fig:simulated_signals}.
The assumed distance is $d=10 \, \mathrm{kpc}$, except
for model s11.2 ($d=5 \, \mathrm{kpc}$). As the low-mass
progenitors u8.1 and z9.6 do not exhibit pronounced spatio-temporal
variations in the neutrino emission, we do not discuss these models
any further in this section.

Different from earlier work on the subject
\citep{marek_09,lund_10,brandt_10,lund_12,tamborra_13}, we not only
study the frequency spectrum for the entire signal or distinct phases
of the evolution, but instead analyze the full time-frequency
structure of the expected IceCube signals by means of a wavelet
analysis. The wavelet transform $\chi(t,p)$ of the rate $\mathfrak{R}$
depends both on time $t$ and on the period $p$, and is given in terms
of the mother wavelet $\psi$ by
\begin{equation}
\label{eq:wavelet}
\chi(t,p)=
\frac{1}{\sqrt{|p|}}
\int\limits_{-\infty}^{\infty} (\mathfrak{R}(t)+\mathfrak{R}_0) \ \psi^\star \left(\frac{t'-t}{p} \right) \ud t' .
\end{equation}
For evaluating Equation~(\ref{eq:wavelet}), we use the discrete
Poisson realization of the binned IceCube data (including
the background), and we employ the
Morlet wavelet as mother wavelet with a scaling parameter\footnote{The
scaling parameter determines the width of the wavelet in terms
of the wavelength. We opt for a relatively small value to achieve
better temporal resolution.} $\sigma=6$:
\begin{eqnarray}
\nonumber
\psi (x) &=& \left(1+e^{-\sigma^2}-2 e^{-3\sigma^2/4}\right)^{-1/2}
\pi^{-1/4} e^{-x^2/2} \\ &&\times \left(e^{i \sigma x} - e^{-\sigma^2/2}
\right).
\end{eqnarray}
 In order to assess the significance of features in the wavelet
  spectrogram, we consider the squared signal-to-noise ratio $(S/N)^2$
  between the absolute square of the wavelet transform and the expectation
  value $\langle |\chi_\mathrm{noise}|^2 \rangle$ due to the
  background,
\begin{equation}
\label{eq:sn}
(S/N)^2=
\frac{|\chi|^2(t,p)}{\langle |\chi_\mathrm{noise}|^2 \rangle}.
\end{equation}
The computation of $\langle |\chi_\mathrm{noise}|^2 \rangle$
for Poissonian noise in different time bins is described
in the Appendix. For the discrete wavelet transform, $\langle |\chi_\mathrm{noise}|^2
\rangle$ depends weakly on the period $p$ as long as $p \gg \Delta t$
and $p \ll T$ (where $T$ is length of the time series). To avoid
spuriously high signal-to-noise ratios for $p \approx \Delta t$, we always
use the noise level at $p \approx 50 \, \mathrm{ms}$. 

Wavelet spectrograms showing $(S/N)^2$ for some observer directions
and different distances are shown in Figures~\ref{fig:wavelet_s25},
\ref{fig:dependence_on_distance} (s25), \ref{fig:wavelet_s15s7b2}
(s15s7b2), \ref{fig:wavelet_s11} (s11.2) and \ref{fig:wavelet_s27}
(s27).

\begin{figure*}
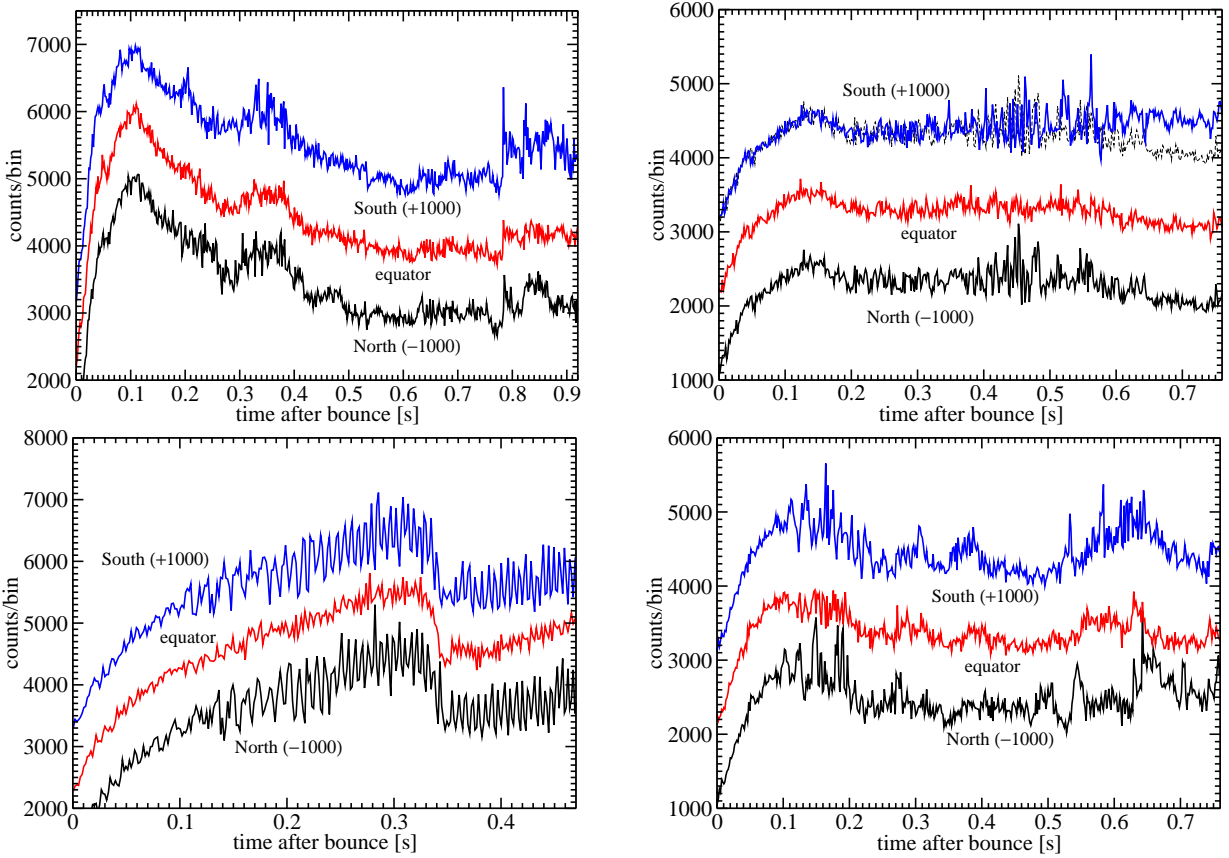

  \plottwo{f10a.eps}{f10b.eps}
  \plottwo{f10c.eps}{f10d.eps}
  \caption{Simulated IceCube signals (including background) for models
    s11.2 (top left, at $5 \, \mathrm{kpc}$), s15s7b2 (top right, at
    $10 \, \mathrm{kpc}$), s25 (bottom left, at $10 \, \mathrm{kpc}$),
    and s27 (bottom right, at $10 \, \mathrm{kpc}$) for observers
    situated in the north/south polar directions and in the equatorial
    plane. Note that we use an offset of $\pm 1000$ for polar
    observers to avoid overlapping curves. For model s15s7b2 (top
    right panel), the count rate in the equatorial plane is shown as
    dashed black line with an offset of $+1000$ to illustrate the
    enhanced emission in the southern hemisphere due to the asymmetric
    explosion geometry.
\label{fig:simulated_signals}}
\end{figure*}

The simulated real-time signals in Figure~\ref{fig:simulated_signals}
show clearly discernible temporal variations on top of the secular
evolution that we already discussed for the total angle-averaged
luminosity. The wavelet spectrograms reveal that these variations are indeed
caused by spatio-temporal modulations in the emission and not by white
noise due the detector background. It is convenient to analyze the
distinct fingerprints of the pre-explosion phase and the explosion
phase separately.

\begin{figure*}
  \plottwo{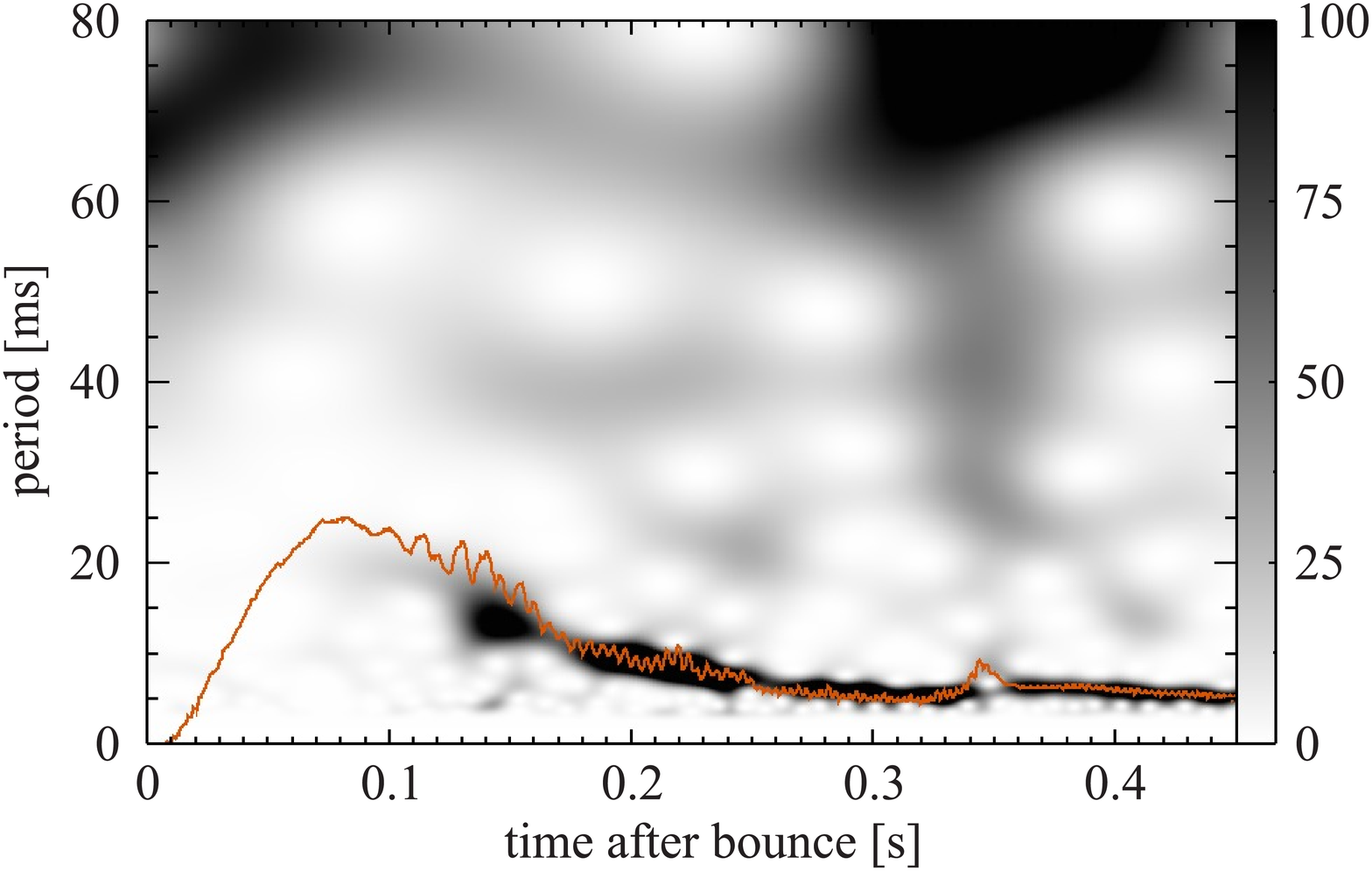}{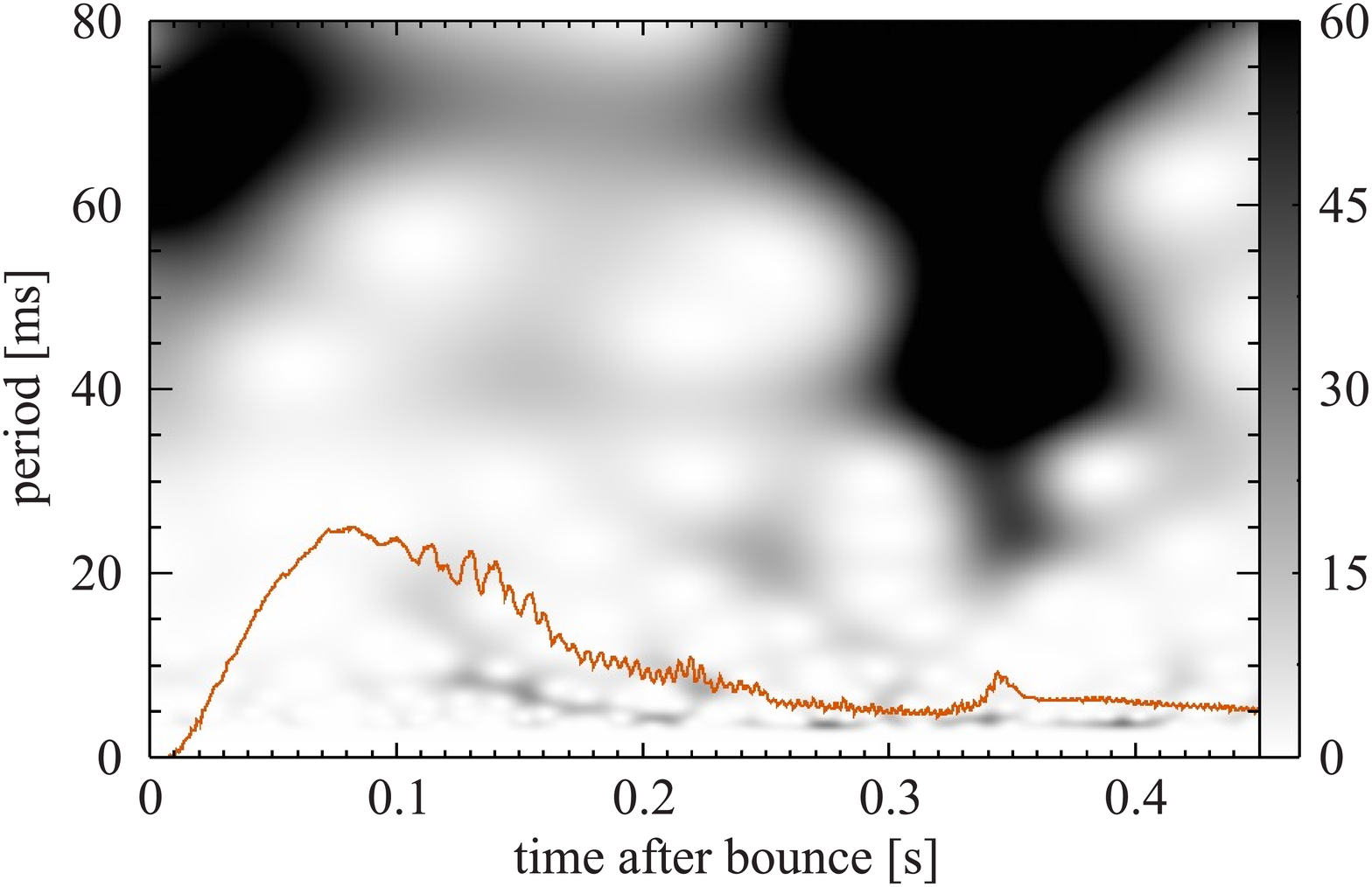}
  \plottwo{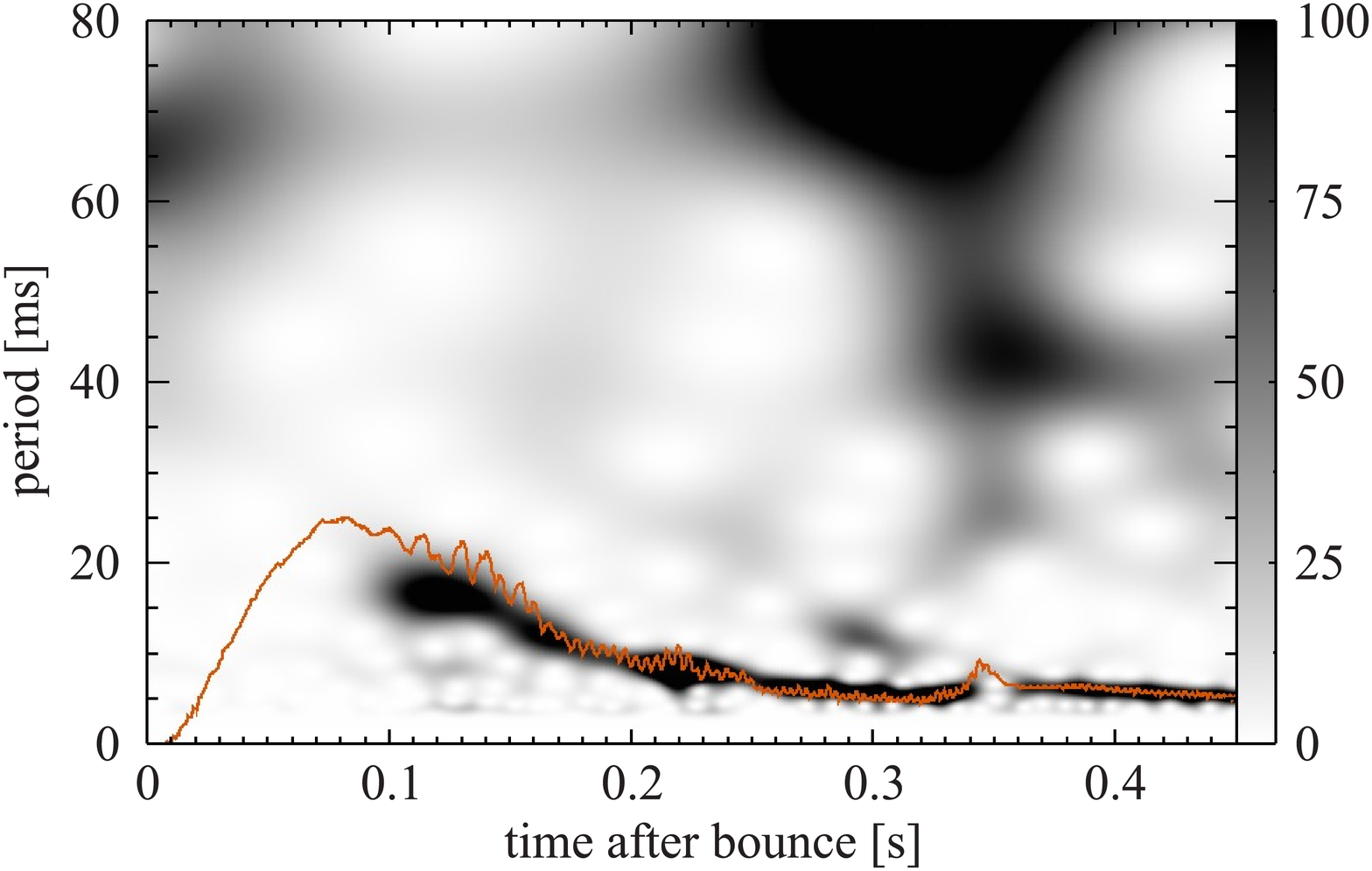}{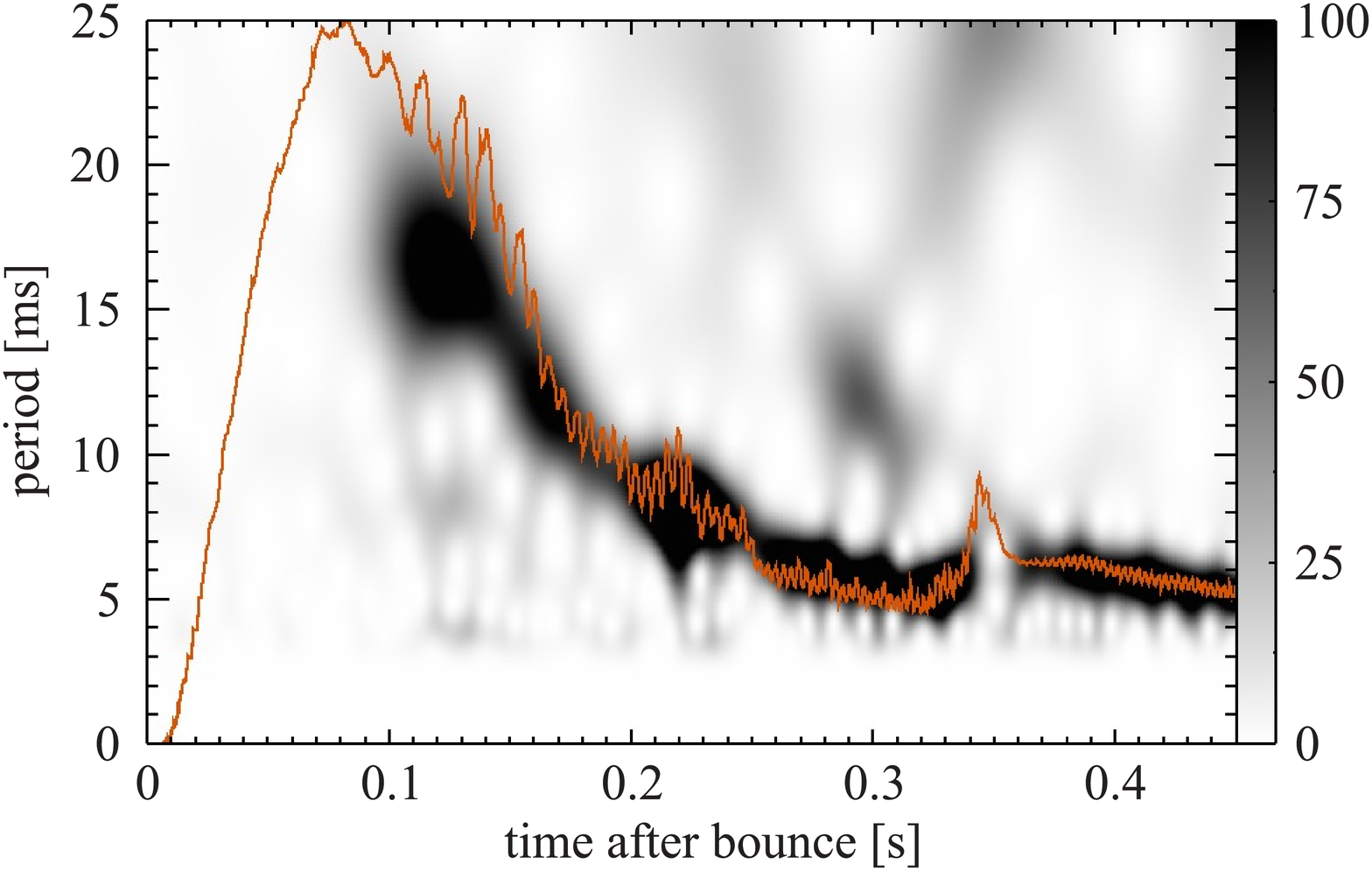}
  \caption{Wavelet spectrograms of simulated IceCube signals for model
    s25 for observers situated at a distance of $10 \, \mathrm{kpc}$ in
    the north polar direction (top left panel), in the equatorial
    plane (top right panel), and along the south polar axis (bottom
    panels).  The SASI period predicted by
    Equation~(\ref{eq:sasi_period}) is shown as a red curve in each
    panel. The colorbar shows the scale for the signal-to-noise
ratio computed according to Equation~(\ref{eq:sn}).
\label{fig:wavelet_s25}
}
\end{figure*}

\begin{figure}
  \plotone{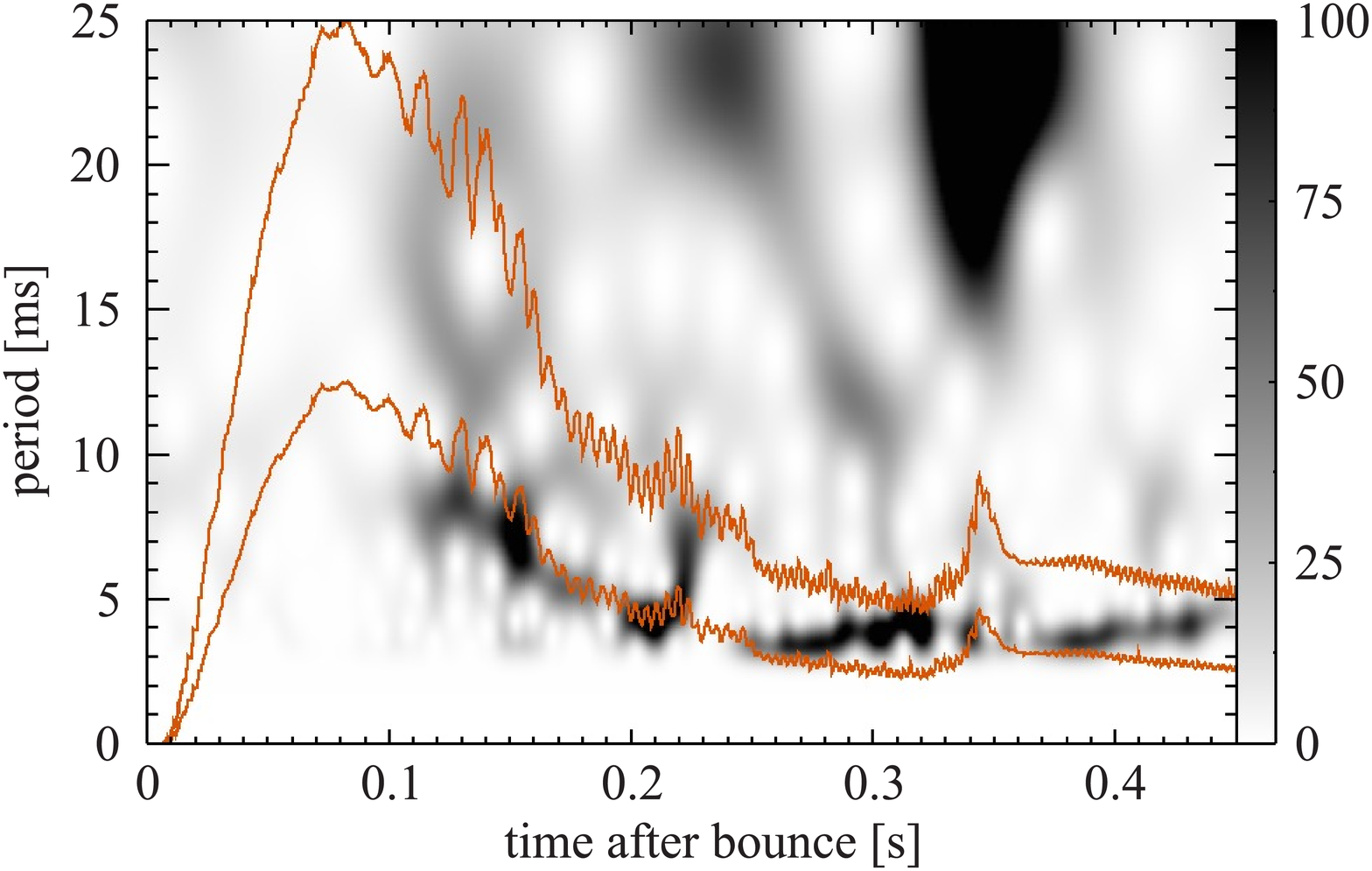}
  \plotone{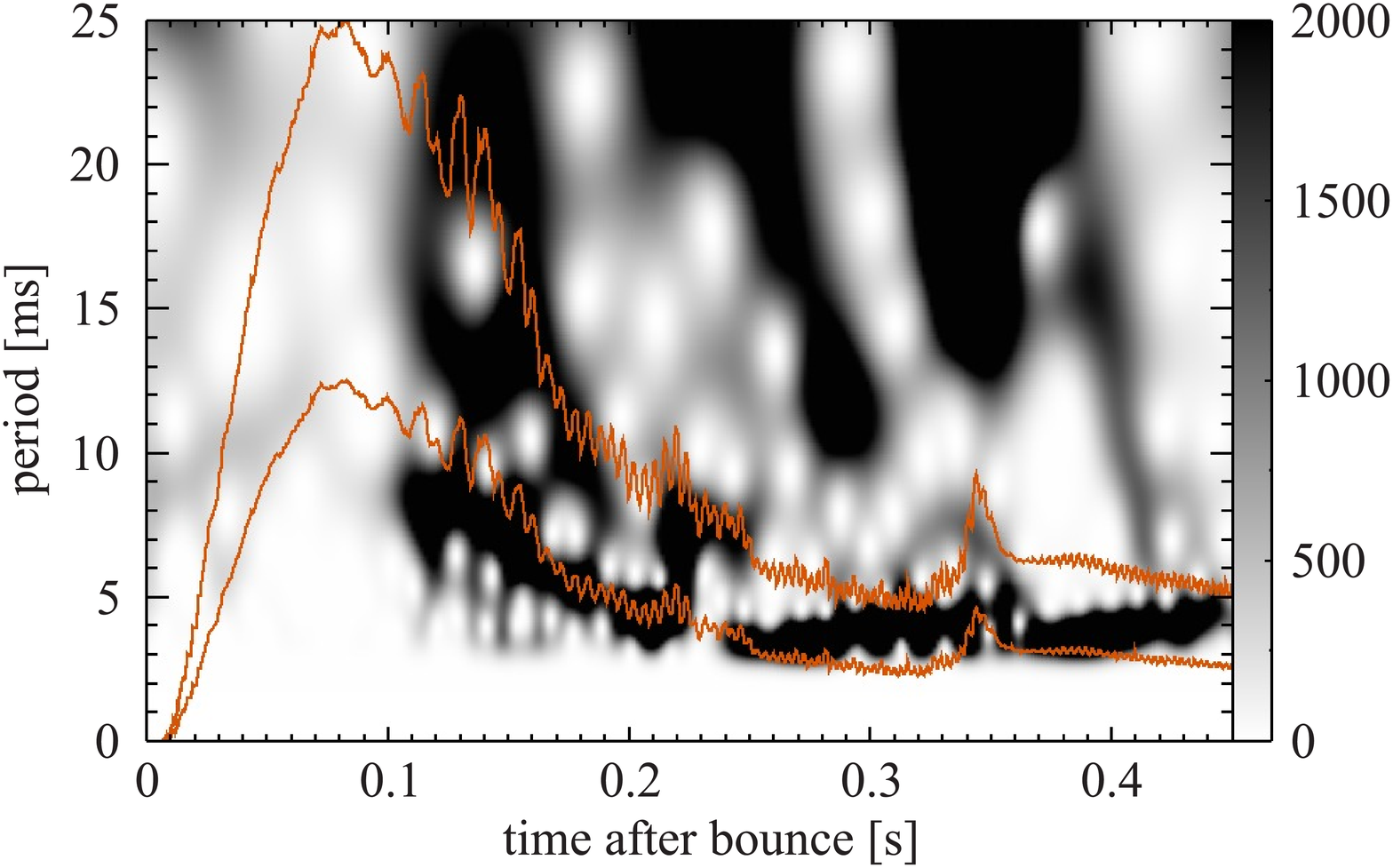}
  \caption{Wavelet spectrograms of simulated IceCube signals for model
    s25 for observers situated in the equatorial plane at a distance
    of $5 \, \mathrm{kpc}$ (top panel) and $2 \, \mathrm{kpc}$ (bottom
    panel). The grayscale indicates the squared signal-to-noise
    ratio $(S/N)^2$ (Equation~\ref{eq:sn}). Red curves show the
    predicted values of $T_\mathrm{SASI}$ and $T_\mathrm{SASI}/2$
    according to Equation~(\ref{eq:sasi_period}). The colorbar shows the scale for the signal-to-noise
ratio computed according to Equation~(\ref{eq:sn}).
\label{fig:dependence_on_distance}}
\end{figure}

\begin{figure}
  \plotone{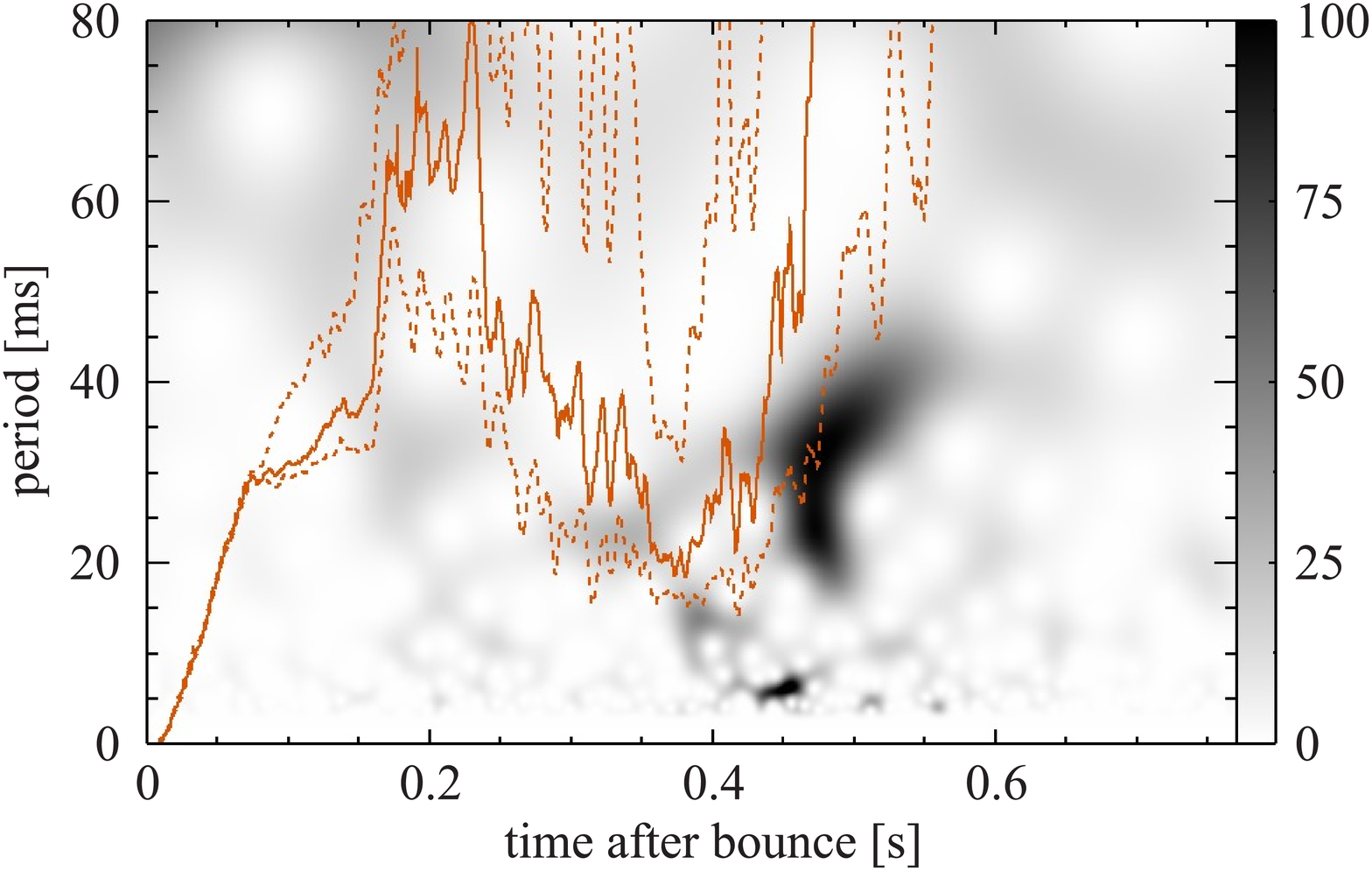}
  \plotone{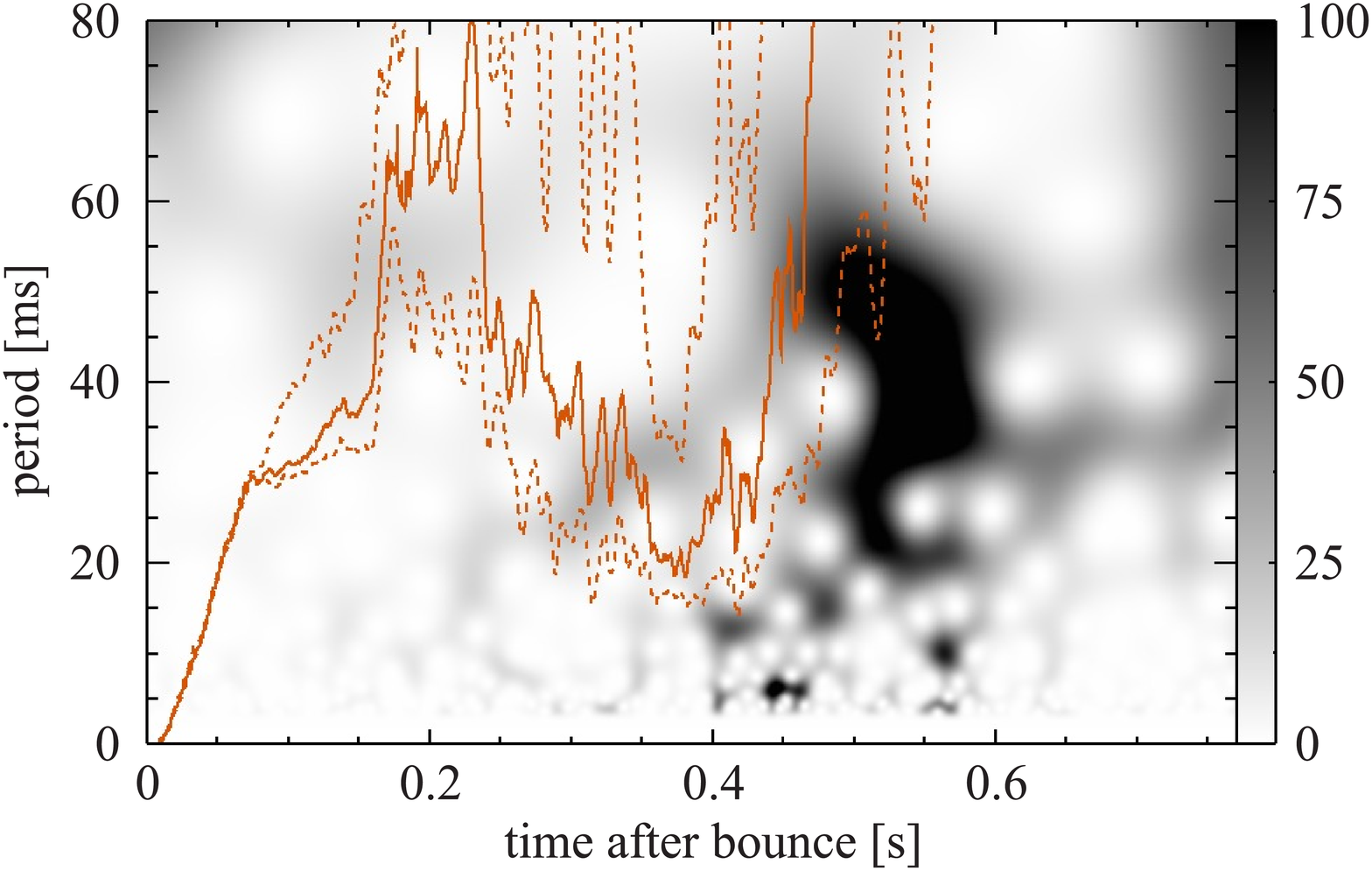}
  \caption{Wavelet spectrograms of simulated IceCube signals for model
    s15s7b2 for observers situated at a distance of $10
    \, \mathrm{kpc}$ in the north polar direction (top panel) and the
    south polar direction (lower panel).  The SASI period predicted
    by Equation~(\ref{eq:sasi_period}) is shown as a red curve in each
    panel. The dashed red lines show the estimates based on
    the maximum and minimum shock radius. The colorbar shows the scale for the signal-to-noise
ratio computed according to Equation~(\ref{eq:sn}).
\label{fig:wavelet_s15s7b2}
}
\end{figure}

\subsection{Signatures of SASI Oscillations}
The most remarkable feature of neutrino emission in the pre-explosion
phase is seen in strongly SASI-dominated models like s25 and s27
(bottom panels of Figure~\ref{fig:simulated_signals}), which exhibit a
strong quasi-periodic modulation of the signal for observers located
along the direction of the SASI sloshing motions
\citep{marek_09,ott_08_a,brandt_10,lund_10,lund_12,tamborra_13}. In a
model like s15s7b2 (top right panel of
Figure~\ref{fig:simulated_signals}), in which the SASI is probably
present but where strong convective overturn also develops, such modulations are still
visible but far less pronounced. The modulations are particularly
strong for a model like s25 that either fails to explode or would
eventually explode at very late times.

In such a case, SASI activity may persist over several hundreds of
milliseconds with little activity of parasitic instabilities on top of
the SASI, and thus with a well-defined periodicity of the
oscillations. This gives rise to a distinct pattern in the wavelet
spectrograms (Figures~\ref{fig:wavelet_s25} and
\ref{fig:dependence_on_distance}) that shows a continuous decrease of
the SASI period as the shock retracts. This decrease may be
interrupted by sudden drops in the accretion rate associated with the
infall of a composition shell interface in the progenitor.  The
changing conditions when such an interface falls through the shock
cause a transient interruption of the quasi-periodic shock
oscillations (Figure~\ref{fig:wavelet_s25}). In the case of model s25,
the drop in the accretion rate is clearly reflected in the
time-dependent neutrino signal (bottom left panel of
Figure~\ref{fig:simulated_signals}).  If SASI activity resumes
afterward, it may do so with a somewhat longer oscillation period than
before, which is reflected as a
small step in the frequency band in the bottom right panel of
Figure~\ref{fig:wavelet_s25}.

The SASI-induced modulations of the neutrino emission are most
pronounced for observers located along the direction of SASI sloshing,
but even for observers viewing the supernova from an orthogonal
direction, the time-frequency structure of the neutrino signal still
shows fingerprints of SASI activity (top right panel of
Figure~\ref{fig:wavelet_s25}), in particular if we assume a distance
smaller than $10 \, \mathrm{kpc}$, as shown by the spectrograms in
Figure~\ref{fig:dependence_on_distance}. However, the frequency
appears to be much broader in this case, reaching to significantly
lower periods. The bottom panel of
Figure~\ref{fig:dependence_on_distance} (which shows the spectrogram
for an observer in the equatorial plane at $2 \, \mathrm{kpc}$)
suggests that there may be two emission bands and that the dominant
modulation frequency is roughly twice as large as for a polar
observer. This is just a reflection of the fact that an excursion of
the shock in either direction will affect the neutrino emission in the
equatorial plane in the same manner. For a real detection, the
relative orientation of the observer to a putative sloshing mode of
the SASI would be unknown, but the presence or absence of a secondary
emission peak at lower frequencies (longer periods) would allow a
discrimination between the ``real'' SASI frequency band and the
artificial ``overtone'' that occurs for an observer not located along
the direction of the sloshing mode.  For a pure spiral
mode, this overtone would probably be absent, and the viewing angle
would only affect the amplitude of the modulations.

The IceCube signal from a Galactic supernova could thus not only provide
unambiguous evidence about SASI activity in the supernova core,
but might even give some hints about the involved modes (sloshing
vs.\ spiral modes) and could allow us to directly follow the temporal
evolution of the SASI frequency. This would also provide qualitative
information about the shock radius and the proto-neutron star radius,
which set the oscillation period of the SASI 
\citep{foglizzo_07,scheck_08}. Quite remarkably, it is possible to
formulate a rather simple quantitative relation between the
time-dependent period of the IceCube signal modulations and these two
radii.  If the SASI is due to an advective-acoustic cycle, its period
is given by the sum of the advective and acoustic time-scales for
perturbations traveling between the (angle-averaged) shock radius
$r_\mathrm{sh}$ and the radius of maximum deceleration $r_\nabla$
\citep{foglizzo_07,scheck_08}:
\begin{equation}
\label{eq:sasi_timescale}
T_\mathrm{SASI}
=
\tau_\mathrm{adv}+\tau_\mathrm{ac}
=
\int_{r_\nabla}^{r_\mathrm{sh}} \frac{\ud r}{|v_r|}+
\int_{r_\nabla}^{r_\mathrm{sh}} \frac{\ud r}{c_s-|v_r|}.
\end{equation}
Here, $v_r$ and $c_s$ are the (average) radial velocity and the
local sound speed. The velocity profile between $r_\mathrm{sh}$
and $r_\nabla$ is roughly linear, 
\begin{equation}
\label{eq:velocity_profile}
v_r \sim -\beta^{-1} \sqrt{\frac{G M}{r_\mathrm{sh}}} \left(\frac{r}{r_\mathrm{sh}}\right),
\end{equation}
where $\beta$ is the ratio between the post-shock and pre-shock density.
Since the flow becomes very subsonic close to $r_\nabla$, the advection
time-scale $\tau_\mathrm{adv}$ will typically be the dominant term
that decides about the scaling of $T_\mathrm{SASI}$ with the
parameters of the accretion flow ($M$, $r_\mathrm{sh}$, $r_\nabla$).
Plugging in Equation~(\ref{eq:velocity_profile}) into
Equation~(\ref{eq:sasi_timescale}) and neglecting the acoustic time-scale
results in
\begin{equation}
\label{eq:sasi_period_analytic}
T_\mathrm{SASI} \propto
r_\mathrm{sh}^{3/2} M^{-1/2} \ln 
\left(\frac{r_\mathrm{sh}}{r_\nabla} \right).
\end{equation}
The radius of maximum deceleration $r_\nabla$ is somewhat difficult to
infer from simulations, and its dependence on other proto-neutron star
parameters (mass, core radius $R_\mathrm{core}$, neutrinosphere
radius, gain radius, surface temperature) is rather complicated since
the location of the coupling region for vorticity perturbations and
acoustic perturbations also depends on the density gradient in the
cooling region \citep{scheck_08}. However, we can formulate an
empirical scaling law in terms of the proto-neutron star radius
$r_\mathrm{PNS}$ (defined by a fiducial density of $10^{11}
\, \mathrm{g} \, \mathrm{cm}^{-3}$ as in
\citealt{mueller_12a,bruenn_13,suwa_13}) for the SASI oscillation
period:
\begin{equation}
\label{eq:sasi_period}
T_\mathrm{SASI} = 19 \, \mathrm{ms}
\left(\frac{r_\mathrm{sh}}{100 \, \mathrm{km}}\right)^{3/2}
\ln 
\left(\frac{r_\mathrm{sh}}{r_\mathrm{PNS}} \right).
\end{equation}
Here, we have ignored the relatively weak dependence on $M$ in
Equation~(\ref{eq:sasi_period_analytic}), which gives a somewhat better fit
to the observed SASI periods when using $r_\mathrm{PNS}$ instead of
$r_\nabla$ to estimate the advection time-scale.  The prediction for
$T_\mathrm{SASI}$ is indicated in Figures~\ref{fig:wavelet_s25},
\ref{fig:wavelet_s15s7b2}, \ref{fig:wavelet_s11} and \ref{fig:wavelet_s27} by a red
curve. In Figure~\ref{fig:dependence_on_distance}, $T_\mathrm{SASI}/2$
is also shown.

While Figures~\ref{fig:wavelet_s25} and
\ref{fig:dependence_on_distance} demonstrate that
Equation~(\ref{eq:sasi_period}) describes models with continuous SASI
activity very well, one should also note that the models s15s7b2
(Figure~\ref{fig:wavelet_s15s7b2}) and s27
(Figure~\ref{fig:wavelet_s27}) show some activity around the expected
frequency during the phases where the SASI is active (prior to the
onset of the explosion at $\sim 125 \, \mathrm{ms}$ in s27) and during
the phase of strong shock retraction around $300\ldots 350
\, \mathrm{ms}$ in s15s7b2). Even the supposedly convective model s11.2
(Figure~\ref{fig:wavelet_s11}) shows some (faint)
broadband activity around the SASI frequency prior to the onset of the
explosion (at $p\approx 35 \, \mathrm{ms}$ at a post-bounce time
of $\approx 100 \, \mathrm{ms}$). Among these exploding models with
somewhat less extended and less regular SASI activity than model s25,
model s27 is particularly noteworthy as it shows clear signs of an
\emph{increasing} shock oscillation period -- and hence an
increasing shock radius -- from $100 \, \mathrm{ms}$ onward. This is
one of the characteristic features of exploding models, which we
discuss in the following section.

However, different from a model like s25, the time-frequency structure
of the neutrino signal of these models is a less robust indicator for
SASI as the physical mechanism underlying the emission modulation. The
broadband nature, the intermittent character of the signal
modulations, and the temporal variability of the dominant frequency
point to a significant and perhaps dominant role of convection in
models like s15s7b2 and s11.2.

\begin{figure}
  \plotone{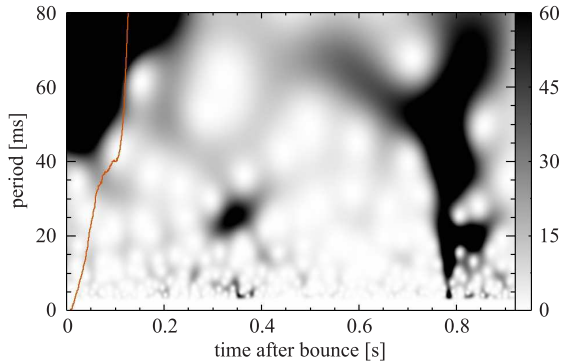}
  \caption{Wavelet spectrogram of the simulated IceCube signal for an
    observer in the south polar direction at $5 \, \mathrm{kpc}$ for
    model s11.2. The SASI period predicted by
    Equation~(\ref{eq:sasi_period}) is shown as a red curve. Note the stripe-like patterns indicating the
    formation of a new downflow onto the proto-neutron star
    around $800 \, \mathrm{ms}$ and also around $350 \, \mathrm{ms}$. The colorbar shows the scale for the signal-to-noise
ratio computed according to Equation~(\ref{eq:sn}).
\label{fig:wavelet_s11}
}
\end{figure}

\begin{figure}
  \plotone{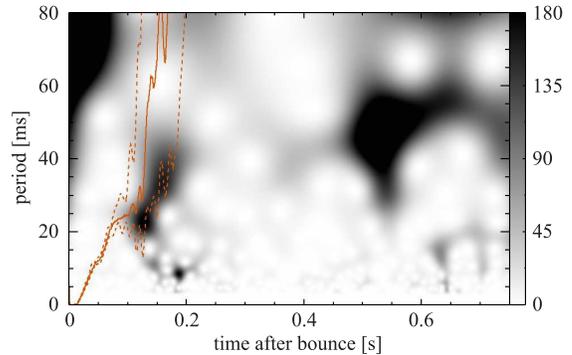}
  \caption{Wavelet spectrogram of the simulated IceCube signal for an
    observer in the north polar direction at $10 \, \mathrm{kpc}$ for
    model s27. The SASI period predicted by
    Equation~(\ref{eq:sasi_period}) is shown as a red curve. The
    dashed red lines show the estimates based on the maximum and
    minimum shock radius.  The spectrogram reveals enhanced
    post-explosion accretion onto the proto-neutron star around $550
    \, \mathrm{ms}$, $650 \, \mathrm{ms}$, and $720 \, \mathrm{ms}$.
The colorbar shows the scale for the signal-to-noise
ratio computed according to Equation~(\ref{eq:sn}).
\label{fig:wavelet_s27}
}
\end{figure}

\begin{figure}
  \plotone{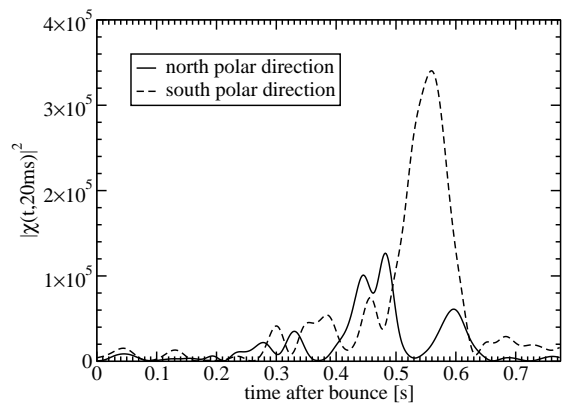}
  \caption{Squared absolute wavelet amplitude $|\chi|^2$ for a period
    of $20 \, \mathrm{ms}$ for model s15s7b2. The solid and dashed
    lines show $|\chi|^2$ for observers situated along the north and
    south polar directions, respectively.
\label{fig:amplitude_20ms_s15s7b2}
}
\end{figure}

\begin{figure*}
\plotone{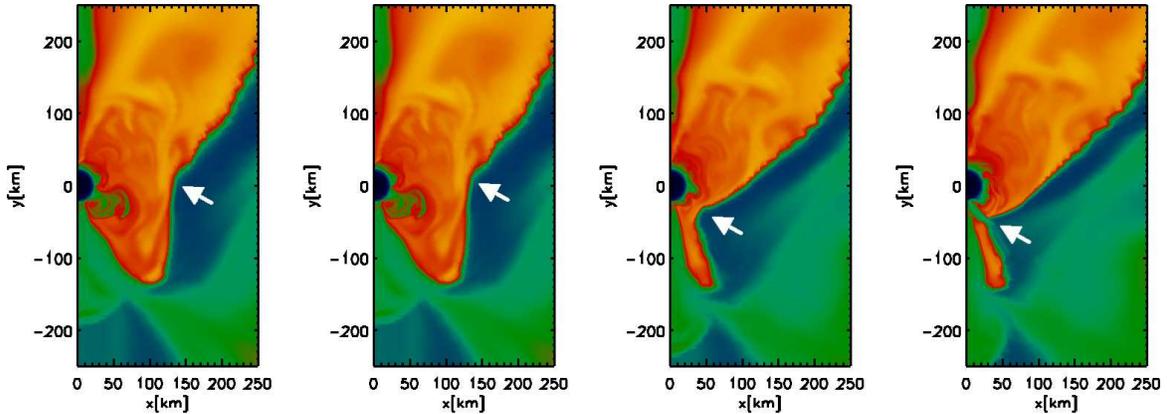}
\caption{Snapshots of the entropy $s$ (ranging
  from $0k_b/\mathrm{nucleon}$ (black) to $35k_b/\mathrm{nucleon}$ (yellow) in
  the region around the proto-neutron star at post-bounce times of
  $812.3 \, \mathrm{ms}$, $816.5 \, \mathrm{ms}$, $819.7 \, \mathrm{ms}$,
  and $821 \, \mathrm{ms}$. The white arrow indicates where cold
  infalling material penetrates into the hot high-entropy bubble
  and eventually forms a new downflow, which then supplies fresh
  material to the cooling region and causes a small, burst-like
  enhancement of the neutrino emission.
\label{fig:downflow_burst}
}
\end{figure*}

\subsection{Signatures of the Explosion Phase}
\label{sec:variations_explosion}
As discussed in Section~\ref{sec:explosion_spherical_average}, the
directionally averaged neutrino fluxes cannot serve as robust
indicators for the onset of the explosion. The relatively abrupt
decline of the luminosity seen in artificial 1D explosion models is
often absent in multi-D simulations. A step-like decline in the
electron neutrino and antineutrino luminosities rather points to the
infall of a composition interface.

Fortunately, the spatio-temporal variations of the neutrino-emission provide
several clues for diagnosing the transition from the accretion phase
to the explosion phase. However, these fingerprints are not
immediately obvious from a superficial visual inspection of the
simulated IceCube detection rates in
Figure~\ref{fig:simulated_signals}.  Fluctuations in the observable
neutrino signal are present both prior to and after shock revival, and the
amplitudes in both phases are not dissimilar. The occurrence of
sustained \emph{global} emission anisotropies (e.g.\ in model s15s7b2,
top right panel of Figure~\ref{fig:simulated_signals}) is a
qualitative difference to the pre-explosion phase. Such global
anisotropies can result from one-sided accretion in very asymmetric
explosions (cf.\ our discussion of model s15s7b2 in
\citealt{mueller_12a}), but they can obviously not be observed
directly.

Nevertheless, the transition to the explosion betrays itself by
quantitative and qualitative changes in the fluctuating neutrino
signal: Due to the expansion of the shock, the typical frequency of
the fluctuations decreases (Figures~\ref{fig:wavelet_s15s7b2} and
\ref{fig:wavelet_s27}), and the fluctuations lack the well-defined
periodicity familiar from SASI-dominated models in the
pre-explosion phase.  In the wavelet spectrograms of the exploding
models s11.2, s15s7b2 and s27 (intermittent) broadband activity for
periods longer than $\gtrsim 20 \, \mathrm{ms}$ therefore dominates
over short-period fluctuations. In our models, a typical period of
fluctuations around $\sim 20 \ldots 25 \, \mathrm{ms}$ seems to provide
a very rough dividing line between non-exploding and exploding
models. Figure~\ref{fig:amplitude_20ms_s15s7b2} illustrates that for
model s15s7b2 the activity peak at $p=20 \, \mathrm{ms}$ coincides very
well with the phase of shock revival. \emph{A neutrino signal
  dominated by fluctuations in the range of $40 \ldots 50
  \, \mathrm{Hz}$ is therefore probably a good indicator for shock
  revival. }

It is worth noting that even during the phase of shock revival, the
spectrum of the emission fluctuations still shows a preferred
frequency at least in some models, albeit that the peak is rather
broad.  Model s15s7b2, for example, shows strong signal fluctuations
around $500 \, \mathrm{ms}$ for an observer along the north polar axis.
The typical period of these fluctuations is in rough agreement with
the advection time-scale between the \emph{minimum} shock radius and
the proto-neutron star (Figure~\ref{fig:wavelet_s15s7b2}
and also Figure~\ref{fig:wavelet_s27} for the corresponding
feature in model s27). This could
suggest that some advective-acoustic feedback mechanism (which may or
may not be identified with the SASI) is still active at this stage and
provides a preferred time-scale for variations in the accretion flow
and the neutrino emission.

Early fallback of some of the shocked material onto the proto-neutron
star also gives rise to characteristic signatures in the neutrino
emission after the onset of the explosion, for which model s11.2 is a
prime example. As already discussed in \citet{mueller_12a,mueller_13},
the energy of the hot, neutrino-heated material is relatively low in
this case so that the high-entropy bubbles fail to push out all the
material swept up by the shock after the onset of the explosion. Much
of the shocked material is therefore channeled onto the proto-neutron
star through long-lived downflows, but on occasion, some of the
swept-up material also penetrates the expanding high-entropy bubbles
to form a new downflow as illustrated in
Figure~\ref{fig:downflow_burst}. Such a new downflow not only excites
oscillations of the proto-neutron star surface that give rise to
burst-like gravitational wave emission \citep{mueller_13}, but also
carries fresh material into the cooling region.

This results in a sudden jump in the neutrino emission, which would be
observable by IceCube for a Galactic supernova at a distance of $5
\, \mathrm{kpc}$ as can be seen in the top left panel of
Figure~\ref{fig:simulated_signals} (with two prominent bursts and
$\sim 780 \, \mathrm{ms}$ and $\sim 820 \, \mathrm{ms}$). Although the
mini-bursts in the neutrino-emission are more strongly pronounced for
an observer in the south polar direction (i.e.\ above the hemisphere
where the downflow develops), they are visible from any direction as
the newly injected material is quickly redistributed across the whole
cooling region. As the expected count rate rises by several hundred,
these events are clearly distinguishable from statistical fluctuations
in the signal, which are of the order of $\sqrt{\mathfrak{R}} \approx
63$ during the relevant phase. If the jump in the detection rate is
statistically significant, such mini-bursts also leave a distinct
trace in the wavelet spectrogram in the form of extended vertical
stripes. These patterns are most clearly visible in the case of model
s11.2 (Figure~\ref{fig:wavelet_s11}).  Model s27 also shows such mini-bursts,
albeit a little less sharp (Figure~\ref{fig:wavelet_s27}),
e.g.\ at $\sim 550 \, \mathrm{ms}$ and $\sim 650 \, \mathrm{ms}$.  The
detection of such signatures would not only indicate that an explosion
has been initiated prior to the burst, but would also indicate that
the explosion is (still) reasonably weak at this stage such that outflow
and accretion downflows persist simultaneously for an extended period of time.

\section{Summary and Conclusions}
\label{sec:summary}
Based on six general relativistic 2D simulations of progenitors
between $8.1 M_\odot$ and $27 M_\odot$, we presented a detailed
analysis of the neutrino emission from core-collapse supernovae from
the bounce to the explosion phase. We discussed both the secular
evolution of the total, angle-integrated neutrino emission, which is
largely regulated by the continuous accretion of material and the
contraction of the proto-neutron star, as well as spatio-temporal
variations in the neutrino emission caused by nonaspherical
hydrodynamic instabilities like convection and the SASI. Using
simulated signals of a future Galactic supernova in IceCube
\citep{abbasi_11,salathe_12}, we studied the observability of these
spatio-temporal variations and showed how detailed information
about the supernova core could be extracted from the observed
neutrino signal by means of a time-frequency analysis.

Our most salient findings can be summarized as follows:
\begin{enumerate}
\item Prior to the onset of the explosion, the evolution of the total
  neutrino flux and the neutrino mean energies is in qualitative
  agreement with recent 1D models of the accretion phase
  \citep{liebendoerfer_03,liebendoerfer_04,buras_06_b,marek_09,fischer_09,oconnor_13}
  although proto-neutron star convection affects the neutrino emission
  on the level of $10 \ldots 20 \%$ \citep{buras_06_b}. Similar to
  other modern neutrino hydrodynamics simulations, and in stark
  contrast to some earlier 1D models from the 1980's and 1990's, our
  2D models are characterized by very similar mean energies of
  electron antineutrinos and $\mu/\tau$ neutrinos.  In all but the
  least massive progenitors ($8.1 M_\odot$ and $9.6 M_\odot$), we even
  observe a crossing of the mean energies already within the first few
  hundreds of milliseconds of post-bounce accretion as a result of a
  temperature inversion near the neutrinosphere.
\item The mass of the proto-neutron star emerges as the single most
  important parameter regulating the secular rise of the mean neutrino
  energies, at least for a given equation of state. For individual
  models, we find that the electron antineutrino mean energy
  $\langle E_{\bar{\nu}_e} \rangle$ scales very well with the proto-neutron star mass
  $M$, 
\begin{equation}
\langle E_{\bar{\nu}_e}\rangle \propto M.
\end{equation}
The proportionality constant is slightly progenitor-dependent. This
scaling is a consequence of the roughly adiabatic stratification of
the hot accretion mantle of the proto-neutron star between nuclear
density and the neutrinosphere and the steepness of the mass-radius
relation between $0.5 M_\odot$ and $2 M_\odot$ for nucleonic equations
of state \citep{lattimer_07,steiner_10}.
\item After shock revival, our 2D models differ considerably from
  artificial 1D explosion models \citep{fischer_10}. We find that
  there is no abrupt drop of the electron neutrino and antineutrino
  luminosity to indicate the onset of the explosion because accretion
  downflows persist for a long time and transport fresh material to
  the cooling region.  The accretion luminosity therefore remains high
  and can even rise transiently due to early fallback.
\item The neutrino signal from a Galactic supernova
  could provide \emph{time-dependent} information about the dynamics
  in the supernova core at least for distances $\lesssim 10
  \, \mathrm{kpc}$ if the survival probability of electron
  antineutrinos is high. In a detector like IceCube, strong SASI
  activity will reveal itself by a strong, narrow-banded modulation of
  the detection rate with a period of $T_\mathrm{SASI}$ that is
  directly related to the average shock radius $r_\mathrm{sh}$ and the
  proto-neutron star radius $r_\mathrm{PNS}$:
\begin{equation}
T_\mathrm{SASI} \approx 19 \, \mathrm{ms}
\left(\frac{r_\mathrm{sh}}{100 \, \mathrm{km}}\right)^{3/2} \ln
\left(\frac{r_\mathrm{sh}}{r_\mathrm{PNS}} \right).
\end{equation}
The amplitude of the signal modulation will depend on the observer
direction, as will the presence of an overtone at
$T_\mathrm{SASI}/2$. For sustained SASI activity, the observed ``SASI
chirp'' signal will directly reveal the expansion or contraction of
the shock. Based on model s25, one might speculate that failing
supernovae with a sufficient delay time to black hole collapse
will generally be accompanied by a such ``SASI neutrino chirp''.
\item A wavelet-based time-frequency analysis of the observed signal
  in IceCube could also help to pinpoint the onset of the
  explosion. We predict that the explosion will be accompanied by a
  shift of the typical modulation frequency of the signal below $40
  \ldots 50 \, \mathrm{Hz}$. Moreover, the early fallback of shocked
  material onto the proto-neutron star through new downflows in
  (initially) weak explosions will lead to a detectable burst-like
  \emph{increase} in the electron antineutrino luminosity. In the
  wavelet spectrogram of the signal, such events would manifest
  themselves as localized vertical stripes stretching over an extended
  frequency range.
\end{enumerate}

The temporal variations in the neutrino signal are obviously an
intriguing diagnostics for the dynamics in the supernova core, but
their robustness could be a potential concern. Our results are based
on axisymmetric models (instead of full 3D simulations), rely on the
assumption of a normal neutrino mass hierarchy and disregard possible
non-linear flavor conversion due to neutrino-neutrino refraction.  In
our opinion, it is mainly the uncertainties in the neutrino physics
that make it difficult to decide about the detectability of
spatio-temporal variations in the neutrino signal. In the worst case
of an inverted mass hierarchy, MSW conversion in the stellar envelope could
lead to a complete swap of $\bar{\nu}_e$ and $\bar{\nu}_\mu$
\citep{dighe_00}. As the contribution of the accretion luminosity is
small for $\nu_\mu$'s and $\nu_\tau$'s, this would reduce the
amplitude of the signal fluctuations by a factor of several
\citep{tamborra_13}. Even under such pessimistic conditions,
periodic fluctuations would remain detectable out to a few
$\mathrm{kpc}$ at least for models with strong SASI activity like s25.
However, neither our most optimistic case nor this pessimistic
scenario might be realized in nature. Neutrino-neutrino refraction
effects could lead to flavor swap for at least for certain neutrino
energies (see \citealt{duan_09,duan_10} for a review) already close to
the neutrinosphere, which would in turn affect the outcome of MSW
flavor conversion in the envelope.  The precise conditions for
non-linear neutrino flavor conversion (e.g.\ the interplay of
neutrino-neutrino refraction with the ordinary matter affect and the
role of a non-axisymmetric neutrino distribution in phase space) are a
matter of active debate
\citep{sawyer_09,sarikas_12a,sarikas_12b,saviano_12,cherry_12,mirizzi_12,raffelt_13a,raffelt_13b}. Moreover,
the simple picture of adiabatic MSW flavor conversion in the envelope
might not hold in the presence of sufficiently strong turbulent
density perturbations in convection shells
\citep{kneller_10,kneller_13,lund_13}. These complications preclude
any precise estimate about the amplitude of temporal fluctuations in
the observed neutrino signal. However, uncertainties in the neutrino
physics are unlikely to render these fluctuations completely
undetectable for nearby supernovae driven by vigorous hydrodynamic
iinstabilities.

Our use of the ray-by-ray-plus method for the neutrino transport
instead of a full multi-angle treatment might also be a concern, but
by reprocessing our ray-by-ray-plus results as described in
Section~\ref{sec:reconstruction}, we ensure that we do not grossly
overestimate the amplitude of the fluctuating neutrino signal.

The restriction of our current models to 2D is probably less of a
concern, especially for clearly SASI-dominated models like s25 and
s27. As demonstrated by \citet{hanke_13}, the SASI can grow no less
vigorously in 3D under appropriate conditions, and the concomitant
emission anisotropies are of similar magnitude as in our 2D models
(see \citealt{tamborra_13}). The viewing-angle dependence of
SASI-induced neutrino flux variations may of course be different in
3D, where the SASI can also develop a spiral mode, and where both the
sloshing axis and the plane of the spiral mode can be time-dependent.

It is evident that the temporal variations of the observable neutrino
signal could prove a powerful diagnostic in the case of Galactic
supernova, revealing much more than the mere presence of the SASI and
an ``average SASI'' frequency as discussed in previous studies
\citep{marek_09,ott_08_a,lund_10,brandt_10,lund_12,tamborra_13}.
Together with time-dependent measurements of the neutrino flux and
mean energy, the temporal variations in the neutrino signal could
potentially provide much more than such overall constraints on the
core mass and compactness as discussed previously in the literature
\citep{bruenn_87,burrows_88,oconnor_13}, especially if gravitational
wave spectra were also available to determine the surface g-mode
frequency $f$ of the proto-neutron star (with $f(t) \propto G
M/(r_\mathrm{PNS}^2 \langle E_{\bar{\nu}_e}\rangle) \left(1-G
M/r_\mathrm{PNS} c^2\right)^{2}$ as shown in \citealt{mueller_13}).
The approximate mass-temperature relation $\langle
E_{\bar{\nu}_e}\rangle \propto M$ and the relation for the SASI chirp
might then allow a (tentative) reconstruction of the parameters $M(t)$
(proto-neutron star mass),
$r_\mathrm{PNS}(t)$ (proto-neutron star radius), and $r_\mathrm{sh}
(t)$ (shock radius) of the accretion flow in the pre-explosion phase.

Given the manifold uncertainties in predicting precise neutrino signal
templates (neutrino flavor conversion, opacities in dense matter,
equation-of-state dependence), it is, of course, obvious that the ``signal
inversion problem'' is bound to remain a serious challenge in neutrino
astronomy that goes far beyond the scope of this paper.  We believe,
however, that the work presented here may provide some useful ideas
for interpreting future observations of core-collapse supernovae in
neutrinos.

\acknowledgements This work was supported by the Deutsche
Forschungsgemeinschaft through the Transregional Collaborative
Research Center SFB/TR 7 ``Gravitational Wave Astronomy'' and the
Cluster of Excellence EXC 153 ``Origin and Structure of the Universe''
(http://www.universe-cluster.de). The computations were performed on
the IBM p690 and the IBM iDataPlex system \emph{hydra} of the Computer
Center Garching (RZG), on the Curie supercomputer of the Grand
\'Equipement National de Calcul Intensif (GENCI) under PRACE grant
RA0796, on the Cray XE6 and the NEC SX-8 at the HLRS in Stuttgart
(within project SuperN), on the JUROPA systems at the John von Neumann
Institute for Computing (NIC) in J\"ulich (through grant HMU092 and
through a DECI-7 project grant), and on the Itasca Cluster of the
Minnesota Supercomputing Institute.

\appendix

\section{Noise Level for the Discrete Wavelet Transform}
In order to compute the expectation value $\langle |\chi(t,p)|^2
\rangle$ of the absolute square of the wavelet transform due to the
background, we consider the discrete version of
Equation~(\ref{eq:wavelet}):
\begin{equation}
\chi(t_l,p_k)=
\frac{1}{\sqrt{|p_k|}}
\sum_{i=1}^N \mathfrak{B}(t_i) \ \psi^\star \left(\frac{t_i-t_l}{p_k} \right) \Delta t.
\end{equation}
Here, $\mathfrak{B}(t_i)$ is a random variable denoting the number of 
background events in the $i$-th time bin.

We may assume without loss of generality that $t_l=0$ and compute
a time-independent expectation value $\langle |\chi(p_k)|^2 \rangle$
for the background. $\langle |\chi(p_k)|^2 \rangle$ is given by
\begin{equation}
\langle |\chi(p_k)|^2 \rangle =
\frac{1}{|p_k|}
\sum_{i=1}^N \sum_{j=1}^N \langle \mathfrak{B}(t_i) \mathfrak{B}(t_j)\rangle
 \ 
 \psi^\star \left(\frac{t_i}{p_k} \right) \psi \left(\frac{t_j}{p_k} \right)
 \Delta t^2.
\end{equation}
For uncorrelated Poissonian noise in all the different time bins,
we have
\begin{equation}
\langle \mathfrak{B}(t_i) \mathfrak{B}(t_j)\rangle=
\left\{
\begin{array}{ll}
\langle \mathfrak{B}(t_i)\rangle \langle \mathfrak{B}(t_j)\rangle = \langle \mathfrak{B}\rangle ^2, & i \neq j, \\
\langle \mathfrak{B}(t_i)^2\rangle=2 \langle \mathfrak{B}\rangle^2, & i = j .
\end{array}
\right.
\end{equation}
Here, $\langle \mathfrak{B} \rangle=\mathfrak{R}_0$ is the
(time-independent) expectation value of $\mathfrak{B}(t_i)$. This
implies that $\langle |\chi(p_k)|^2 \rangle$ is given by
\begin{eqnarray}
\langle |\chi(p_k)|^2 \rangle
&=&
\frac{\Delta t^2 }{|p_k|}
\sum_{i=1}^N \sum_{j=1}^N \mathfrak{R}_0^2 (1+\delta_{ij})
 \ 
 \psi^\star \left(\frac{t_i}{p_k} \right) \psi \left(\frac{t_j}{p_k} \right)
\\
\nonumber
&=&
\frac{\Delta t^2}{|p_k|}
\left[
\sum_{i=1}^N  \mathfrak{R}_0^2
 \ 
 \left|\psi \left(\frac{t_i}{p_k} \right)\right|^2
+
\sum_{i=1}^N \sum_{j=1}^N \mathfrak{R}_0^2
 \ 
 \psi^\star \left(\frac{t_i}{p_k} \right) \psi \left(\frac{t_j}{p_k} \right)
\right]
\\
\nonumber
&=&
\frac{\Delta t^2}{|p_k|}
\left[
\sum_{i=1}^N  \mathfrak{R}_0^2 
 \ 
 \left|\psi \left(\frac{t_i}{p_k} \right)\right|^2
+
\mathfrak{R}_0^2 \left|\sum_{i=1}^N \psi \left(\frac{t_i}{p_k} \right) \right|^2
\right]
\\
\nonumber
&=&
\frac{\mathfrak{R}_0^2  \Delta t^2}{|p_k|}
\left[
\sum_{i=1}^N 
 \ 
 \left|\psi \left(\frac{t_i}{p_k} \right)\right|^2
+
 \left|\sum_{i=1}^N \psi \left(\frac{t_i}{p_k} \right) \right|^2
\right].
\end{eqnarray}

\end{document}